\documentclass[%
 twocolumn,
 reprint,
showpacs,
 amsmath,amssymb,
 aps,
prb,
]{revtex4}

\usepackage{graphicx}
\usepackage{dcolumn}
\usepackage{bm}
\usepackage{ulem}
\usepackage{color}
\usepackage{braket}

\begin{document}
\definecolor{bright_red}{cmyk}{0.0,0.55,0.54,0.0}
\definecolor{metallic_blue}{cmyk}{0.7,0.4,0.0,0.5}
\newcommand{\tb}[1]{\textcolor{black}{#1}}
\newcommand{\tm}[1]{\textcolor{black}{#1}}
\newcommand{\tms}[1]{\textcolor{black}{\sout{#1}}}
\newcommand{\tr}[1]{\textcolor{black}{#1}}
\newcommand{\tbr}[1]{\textcolor{black}{#1}}
\newcommand{\tmb}[1]{\textcolor{black}{#1}}
\newcommand{\cyan}[1]{\textcolor{black}{#1}}
\newcommand{\scyan}[1]{\textcolor{black}{\sout{#1}}}
\newcommand{\green}[1]{\textcolor{black}{#1}}

\preprint{XXXX}

\title{Clues and criteria for designing Kitaev spin liquid revealed by\\
\cyan{thermal and spin}
excitations of
honeycomb iridates Na$_2$IrO$_3$ 
}

\author{
 Youhei Yamaji$^1$, Takafumi Suzuki$^2$, Takuto Yamada$^2$, Sei-ichiro Suga$^2$,
Naoki Kawashima$^3$, and Masatoshi Imada$^4$
}
\affiliation{$^1$Quantum-Phase Electronics Center, The University of Tokyo, Tokyo, 113-8656, Japan}
\affiliation{$^2$Graduate School of Engineering, University of Hyogo, Hyogo, Himeji, 670-2280, Japan}
\affiliation{$^3$Institute for Solid State Physics, The University of Tokyo, Chiba, 277-8581, Japan}
\affiliation{$^4$Department of Applied Physics, The University of Tokyo, Tokyo, 113-8656, Japan}

\date{\today}

\begin{abstract}
Contrary to the original expectation, Na$_2$IrO$_3$ is not a Kitaev's quantum spin liquid (QSL) but shows a zig-zag-type antiferromagnetic order in experiments.
Here we propose experimental clues and criteria to measure
how \tm{a} material in hand is close to the Kitaev's QSL state.
For this purpose, we systematically study
thermal and spin
excitations of a generalized Kitaev-Heisenberg model \textcolor{black}{studied by Chaloupka {\it et al}. in Phys. Rev. Lett. {\bf 110},
097204 (2013)} and 
an effective  {\it ab initio} Hamiltonian for Na$_2$IrO$_3$ proposed \tm{by Yamaji {\it et al.}} in Phys. Rev. Lett. {\bf 113}, 107201 (2014), by employing a numerical diagonalization method.
We reveal that \tm{closeness to} the Kitaev's QSL is characterized by the following properties\tbr{,
besides trivial criteria such as reduction of magnetic ordered moments and N\'eel temperatures}:
\tbr{(1)} Two peaks in the temperature dependence of specific heat
at $T_{\ell}$ and $T_h$ caused by the fractionalization of spin to two types of Majorana fermions.
\tbr{(2)} In between the double peak,
prominent plateau or shoulder  pinned at
$\frac{R}{2}\ln 2$ in the temperature dependence of entropy, where $R$ is the gas constant.
\tbr{(3)}
Failure of the linear spin wave approximation at the low-lying excitations of dynamical structure factors.
\tbr{(4)} \tm{Small ratio $T_{\ell}/T_h$ close to or less than 0.03.}
\tm{According to the proposed criteria, Na$_2$IrO$_3$ is categorized to a compound close to the Kitaev's QSL, and is proven to be a promising candidate for the realization of the QSL if the relevant material parameters can further be tuned \tbr{by
making thin film of Na$_2$IrO$_3$ on various substrates or applying axial pressure perpendicular to the honeycomb networks of iridium ions}.
\tbr{Applications of these characterization to (Na$_{1-x}$Li$_x$)$_2$IrO$_3$ and other related materials are also discussed.}}
%
%
%
%
\end{abstract}

\pacs{75.10.Jm, 75.10.Kt, 75.40.Gb, 75.40.-s}
\maketitle
%
%
%

\newcommand{\eqsa}[1]{\begin{eqnarray} #1 \end{eqnarray}}
\section{Introduction}
\label{sec:Introduction}
Enormous efforts to realize quantum spin liquids (QSLs) have been made since the pioneering work of Anderson and Fazekas~\cite{PWAnderson1973,PFazekas1973}.
Geometrically frustrated interactions, for instance antiferromagnetic Heisenberg interactions on a variance of triangular lattice,
have been studied as one promising way to realize the QSL state.
Recently, the Kitaev model on a honeycomb structure~\cite{AKitaev2006} has attracted attention, 
because the ground state is exactly proven to be in a QSL phase.
In the Kitaev model, two spins on the nearest neighbor sites $i$, and $j$ interact by the Ising-type interaction,
\cyan{$K_{x}S^x_iS^x_j$, $K_{y}S^y_iS^y_j$, and $K_{z}S^z_iS^z_j$}, 
where the Ising anisotropy axis depends on the three different bonding directions that compose the honeycomb structure.
These anisotropic interactions cause a strong frustrated effect, distinct from the typical geometrical frustration.

The QSL state in the Kitaev model contains two distinct phases, namely \tm{ QSLs with gapless excitations from the ground state and those with only gapful excitations~\cite{AKitaev2006}.}
The gapless QSL appears, when the system is located around the symmetric point
where the bond-depending interactions have the same magnitude, \cyan{$K_x=K_y=K_z$}.
If one coupling constant becomes much larger than the other ones,
for example 
\cyan{$|K_z| \gg |K_x|=|K_y|$}
, the gapped QSL state is stabilized.
Interestingly, the both QSL states,
\tm{regardless of whether the excitation is gapped or gapless}, can be described by noninteracting Majorana fermions propagating 
in the background of static $Z_2$ gauge fields ($0$ or $\pi$ flux) that are also written by the localized Majorana fermions~\cite{GBaskaran2007}.
Reflecting the fractionalization of quantum spins into the Majorna fermions,
 the spin correlations for further neighbor sites become
exactly zero in the Kitaev's QSL states, while those for the nearest neighbor survive.

In addition to the ground state properties, low-energy excitations~\cite{GBaskaran2007} and dynamics~\cite{JKnolle2014} of both gapped/gapless QSL states 
have also been investigated by analytical methods. 
The ground state is in the $0$ flux sector and the lowest excitation by a spin flip from the Kitaev's QSL state 
is expressed by adding a `localized' $\pi$-flux pair accompanying an itinerant Majorana fermion.
Since the excited $\pi$-flux pair does not propagate in the Kitaev model, non dispersive mode appears in the low-lying excitation, \tm{where the excitation energy is nonzero}. 
Such a gapped excitation is confirmed in the dynamical spin structure factor (DSF) that can be observed in inelastic-neutron-scattering experiments. 
In the gapped QSL phase, a $\delta$-function peak \tm{emerges at the lowest excitation energy} of the DSF, 
while a sharp peak with a long tail \tm{generating gapless excitations} derived from the incoherent part is observed in the gapless QSL phase.

Quite recently, thermal properties for the Kitaev model on the honeycomb structure 
have been investigated by the quantum Monte Carlo calculations~\cite{JNasu2015}.
The Kitaev model always exhibits a two-peak structure in the specific heat,
associated with the fractionalization of a single quantum spin into
two types of Majorana fermions; one is the itinerant (dispersive) 
Majorana fermion and the other is the localized (dispersionless) Majorana fermion.
The two peaks represent the two crossover temperatures 
associated with the growth of short-range spin correlations (or thermal excitations of the itinerant Majorana fermions) around the high temperature peak 
and freezing of flux (or the thermal excitation of the localized Majorana fermions) around the low-temperature peak, respectively. 
The above characteristic features and pictures for the Kitaev's QSL states are expected to be robust against small perturbations such as magnetic fields and Heisenberg-type interactions.
However,
details of the stability have not been well understood yet for real materials.

In the search for realization of the \green{Kitaev's} QSL states, 
${\rm Na_2IrO_3}$ has attracted attention as one candidate material.  
In ${\rm Na_2IrO_3}$, ${\rm Ir^{4+}}$ ion can be expressed as
\tm{a pseudospin} with the total angular momentum one-half \cite{GJackeli2009}.
We call this \tm{pseudospin} just as `spin' hereafter.  
${\rm IrO_6}$ octahedrons in ${\rm Na_2IrO_3}$ \tm{are} built into a planar structure parallel to the $ab$ plane and 
${\rm Ir^{4+}}$ ions \tm{constitute} a honeycomb structure~\cite{GJackeli2009,JChaloupka2010}.  
Furthermore, the neighboring ${\rm IrO_6}$ octahedrons are connected by sharing two oxygen atoms on an edge.
The two oxygen atoms make bridges connecting the neighboring Ir atoms and both Ir-O-Ir bond angles \tm{are} nearly equal to $90^{\circ}$. 
Because of the Ir-O-Ir \tm{bonds},
the perturbative process generates three different anisotropic interactions between ${\rm Ir^{4+}}$ ions depending on the Ir-Ir bond directions. 
In addition, ${\rm Ir^{4+}}$ ions interact via direct overlap of their orbitals, generating the Heisenberg-type interaction as well. 
Thus, both of the Kitaev-type and Heisenberg-type interactions emerge between the neighboring
${\rm Ir^{4+}}$ ions \cite{JChaloupka2013}, leading to
the Kitaev-Heisenberg model.

Contrary to the initial predictions~\cite{GJackeli2009,JChaloupka2010},
${\rm Na_2IrO_3}$ undergoes a magnetic phase transition to a zigzag antiferromagnetic order at $T_{N} \sim 15  {\rm K}$ \cite{SKChoi2012,FYe2012}. 
In order to \tm{understand} the zigzag ordering, several effective models have been proposed and examined so far~\cite{IKimchi2011,SKChoi2012,JChaloupka2013,VMKatukuri2014,YYamaji2014,YPSizyuk2014}.  
Some of them have succeeded in explaining the thermodynamic quantities such as the specific heat and/or the magnetic susceptibility~\cite{IKimchi2011,JChaloupka2013,YYamaji2014}.

Several {\it ab initio} derivations of effective spin Hamiltonians for Na$_2$IrO$_3$ have shown
that low-energy physics of Na$_2$IrO$_3$ is \tm{roughly} described by dominant Kitaev's Ising-type exchange interactions,
\cyan{$K_x=K_y\gtrsim K_z \simeq -30$ meV}~\cite{YYamaji2014},
while other much smaller interactions including the Heisenberg exchange \tm{eventually hold the key for driving} the zigzag order.
Therefore, one intriguing challenge is to figure out a guideline for materials design
that enables the Kitaev's QSL, \tm{against the small interactions through control of them} by starting with Na$_2$IrO$_3$.

In this paper, we \tm{first} report the characterization of Na$_2$IrO$_3$ based on our numerical studies.
Focusing on {\it ab initio} effective Hamiltonians for Na$_2$IrO$_3$, we evaluate how close ${\rm Na_2IrO_3}$ is located from the Kitaev's QSL phase by introducing several criteria.
\tm{For the effective Hamiltonian of Na$_2$IrO$_3$}, we employ a simple generalized Kitaev-Heisenberg Hamiltonian proposed by Chaloupka, Jackeli, and Khaliullin, in Ref.~\onlinecite{JChaloupka2013}
and an {\it ab initio} effective Hamiltonian proposed in Ref.~\onlinecite{YYamaji2014}.
In the previous works~\cite{YYamaji2014,TSuzuki1_2015},
we discussed the accuracy of these effective models and concluded that the {\it ab initio} model proposed in Ref.~\onlinecite{YYamaji2014} explains not only thermal properties but also dynamics of this compound. 

By comparing the \tm{temperature dependences of specific heat $C$ and the equal-time spin correlations as well as}
the dispersion of the spin excitation based on the linear spin wave approximation and the dynamical spin structure factors $S({\boldsymbol Q},\omega)$,
\tm{we identify three distinct characteristic regions in the phase diagram of the effective Hamiltonians: In addition to the spin liquid phase, the magnetically ordered phase is classified to two distinct regions.
Within the magnetically ordered phases of the generalized Kitaev-Heisenberg model, the system is classified to the category I,
when the quantum spin system shows a single peak structure in the temperature dependence of $C$.
If a quantum spin system shows a two-peak structure in $C$ despite its magnetic order, the system is classified to the category II.
When the ground state is the Kitaev's QSL, the system is classified to the third category, namely, the category III.
As summarized in Table \ref{table:category}, $C$ for the systems in the category III have two-peak structure commonly to the category II.}

\cyan{As shown later, in the category I,}
the low-lying excitations \cyan{of the generalized Kitaev-Heisenberg model} in $S({\boldsymbol Q},\omega)$, which is induced by flipping a spin, 
are well described by using the conventional linear spin wave theory, i.e.,
successfully interpreted as dispersion of (nearly) free magnons.
In contrast to the category I, a system categorized as the category II shows that 
the low-lying excitations in $S({\boldsymbol Q},\omega)$ are not captured by the linear spin wave theory \tm{on a qualitative level}.
The breakdown of the linear spin wave theory is a common property in both categories II and III, 
although the spin wave analysis has been employed in comparing effective Hamiltonians with
experimental results of $A_2$IrO3 ($A$=Na, Li)~\cite{RauKee14,PhysRevB.92.024413} and another Kitaev's QSL candidate $\alpha$-RuCl$_3$~\cite{banerjee2015proximate}.
The above \tm{different categories}
\tm{are not necessarily separated by} the phase transition:
\tm{Categories I and II are separated from III by a quantum phase transition, 
while the states in the categories I and II can be connected smoothly.} 
\tm{All of highly generalized Kitaev models treated in this paper can be represented by one of these three empirical categories consistently in all the physical quantities studied.}
Thus, the spin excitation spectra would \tm{provide us with useful supplementary data for the classification}.
%

We \tm{then} examine the nature of the ground state of the {\it ab initio} Hamiltonian of Na$_2$IrO$_3$ in light of our proposed categorization.
We show that the {\it ab initio} Hamiltonian of Na$_2$IrO$_3$ belongs to the category II.
\tm{This supports that} a better chance of material design \tm{to realize}
the Kitaev's QSL \tm{may exist through a realistic tuning of the material parameters of Na$_2$IrO$_3$.}
In this context, we propose that the temperature dependence of the entropy
\tm{as well as the energy scales measured by the} ratio of the higher- and lower-temperature peaks in $C$
\tm{give a quantitative measure of the} closeness to the Kitaev's QSLs in the category II. 
We believe that this proposal offers a guideline for the materials design.


\begin{table*}
\caption{
Three categories characterized by thermal and magnetic excitations;
peak structures in the temperature dependences of the specific heat $C$ and nature of quasiparticles (QP).
The nature of QP can be discussed from the comparison 
between dynamical spin structure factors $S({\boldsymbol Q},\omega)$ and linear spin wave theory (SW).
}
\begin{ruledtabular}
\begin{tabular}{cllll}
     & $C$ & LRO & QP & $S({\boldsymbol Q},\omega)$ vs. SW \\
\hline
I.   & single peak & magnetic & free magnon & consistent \\ 
II.  & two \tm{peaks} & magnetic & correlated magnon & discrepant \\ 
III. & two \tm{peaks} & no & Majorana & inconsistent \\ 
\end{tabular}
\end{ruledtabular}
\label{table:category}
\end{table*}

\label{sec:Model and Method}
\section{Model and Method}
%
%
\begin{figure*}[htb]
  \begin{center}
  \vspace{0.5cm}
     \includegraphics[width=0.8\linewidth]{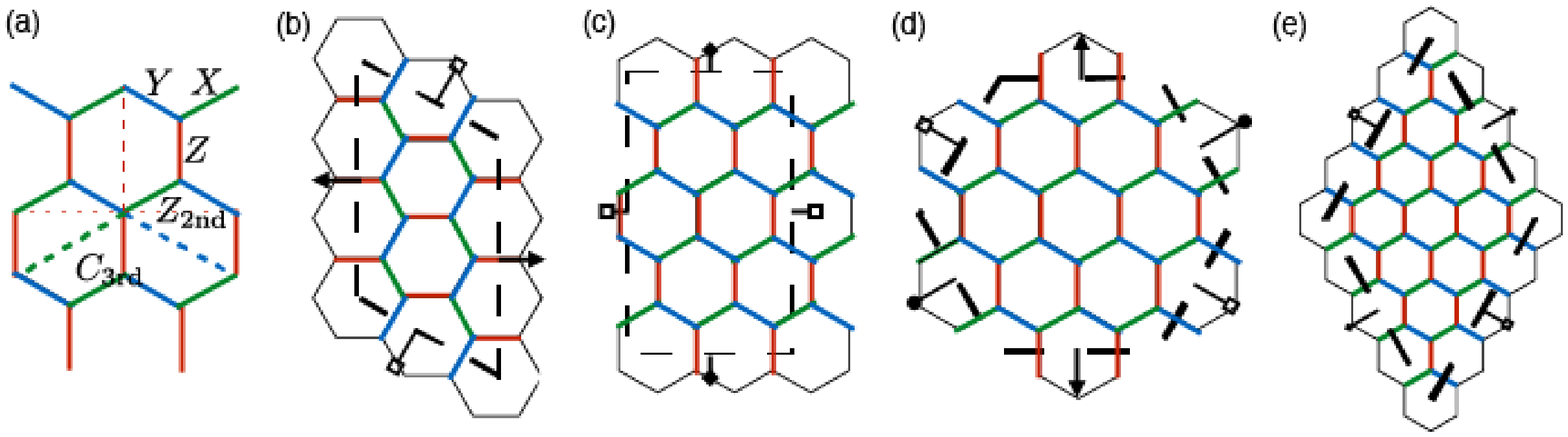}
\caption{\label{fig_Lattice} (Color online) (a) Honeycomb structure model.
Red, blue, and green lines denote $Z$-, $Y$-, and $X$- bond, respectively. 
Dotted (dashed) lines represent the second (third) neighbor bonds.
For the second neighbor bonds, only the bond perpendicular to $Z$-bond, namely, $Z_{\rm 2nd}$-bond is shown because the amplitude for the second neighbor interactions in the {\it ab initio} Hamiltonian is quite small and can be ignored~\cite{YYamaji2014}.
Here, $C_{\rm 3rd}$ denotes the set of the third neighbor bonds. (b-e) $N$=12, 16, 24, and 32-site clusters, respectively.}
 \end{center}
\end{figure*}
%
%
\subsection{Effective Hamiltonian}
\label{subsection:EH}

In this paper, we examine magnetic and thermal excitations of the generalized Kitaev-Heisenberg model
proposed in Ref.~\onlinecite{JChaloupka2013} and the {\it ab initio} Hamiltonian of Na$_2$IrO$_3$
proposed in Ref.~\onlinecite{YYamaji2014}.

\subsubsection{Generalized Kitaev-Heisenberg model}
The generalized Kitaev-Heisenberg model is one of the simplest
models that describe both Kitaev's QSL and zigzag magnetic orders.
The model is parameterized by two exchange couplings, namely the Kitaev-type coupling
\tm{$K=2A\sin \varphi$ and the Heisenberg coupling $J=A\cos \varphi$}, as
\eqsa{
\hat{H}_{\rm CJK}
= 
\sum_{\Gamma=X,Y,Z}
\sum_{\langle i,j \rangle\in \Gamma}
\vec{\hat{S}}_{i}^{T}\mathcal{J}_{\Gamma}\vec{\hat{S}}_j,
}
where \tm{$A$ has the dimension of energy and} $\vec{\hat{S}}_i$ is an SU(2) spin operators $(\hat{S}_i^{x},\hat{S}_i^{y},\hat{S}_i^{z})$ at the $i$-th site, and the matrices of the exchange couplings
for the three nearest-neighbor bonds, $X$-, $Y$-, and $Z$-bond (see Fig.~\ref{fig_Lattice}(a)), are defined as
\eqsa{
\mathcal{J}_{X}
=
\left[
\begin{array}{ccc}
K+J  & 0 & 0 \\
0    & J & 0 \\
0    & 0 & J \\
\end{array}
\right],
}
\eqsa{
\mathcal{J}_{Y}
=
\left[
\begin{array}{ccc}
J  &   0 & 0 \\
0  & K+J & 0 \\
0  &   0 & J \\
\end{array}
\right],
}
and
\eqsa{
\mathcal{J}_{Z}
=
\left[
\begin{array}{ccc}
J & 0 &   0 \\
0 & J &   0 \\
0 & 0 & K+J \\
\end{array}
\right].
}

The ground states of the generalized Kitaev-Heisenberg model range from the Kitaev's QSLs
to trivial magnetically ordered states depending on the control parameter $\varphi$ for fixed $A$ taken positive.
For $\varphi\sim 90^{\circ}$ and $\varphi\sim 270^{\circ}$, the Kitaev's QSLs appear.
When the system size $N=24$ is considered,
the stripy, N\'eel, zigzag, and ferromagnetic orders appear
for $-76,1^{\circ} \lesssim \varphi \lesssim -33.8^{\circ}$,
$-33.8^{\circ} \lesssim \varphi \lesssim 87.7^{\circ}$,
$92.2^{\circ} \lesssim \varphi \lesssim 161.8^{\circ}$,
and $161.8^{\circ} \lesssim \varphi \lesssim 251.8^{\circ}$,
respectively\tbr{, determined
in Ref.\onlinecite{JChaloupka2013}
by exact diagonalization with $N=24$.}

\subsubsection{{\it Ab initio} Hamiltonian of Na$_2$IrO$_3$}

Let us consider a highly generalized form of the Kitaev-Heisenberg model on the honeycomb structure
for the purpose of bridging to the realistic and {\it ab initio} Hamiltonian.
The Hamiltonian is given as
\begin{eqnarray}
\hat{H}_{\lambda}
=
\sum_{\Gamma=X,Y,Z,Z_{\rm 2nd},C_{\rm 3rd}}
\sum_{\langle i,j \rangle\in \Gamma}
\vec{\hat{S}}_{i}^{T}\mathcal{J}_{\Gamma}(\lambda)\vec{\hat{S}}_j,
\label{ham:abinitio}
\end{eqnarray}
where the matrices of the exchange couplings for the three different nearest-neighbor ($X$-, $Y$-, and $Z$-bonds),
the second neighbor ($Z_{\rm 2nd}$-bond), and the third neighbor ($C_{\rm 3rd}$-bond) bonds are given as
\eqsa{
\mathcal{J}_X (\lambda) =
\left[
\begin{array}{ccc}
K'  & 0 & 0 \\
0 & 0 & 0 \\
0 & 0 & 0 \\
\end{array}
\right]
+
\lambda
\left[
\begin{array}{ccc}
0 & I_{2}'' & I_{2}' \\
I_{2}'' & J''  & I_{1}' \\
I_{2}' & I_{1}' & J' \\
\end{array}
\right],
}
\eqsa{
\mathcal{J}_Y (\lambda) =
\left[
\begin{array}{ccc}
0 & 0 & 0 \\
0 & K' & 0 \\
0 & 0 & 0 \\
\end{array}
\right]
+\lambda
\left[
\begin{array}{ccc}
J'' & I_{2}'' & I_{1}' \\
I_{2}'' & 0 & I_{2}' \\
I_{1}' & I_{2}' & J' \\
\end{array}
\right],
}
\eqsa{
\mathcal{J}_Z (\lambda) =
\left[
 \begin{array}{ccc}
0  & 0 & 0 \\
0 & 0 & 0 \\
0 & 0 & K \\
\end{array}
\right]
+
\lambda
\left[
 \begin{array}{ccc}
J   & I_1 & I_2 \\
I_1 & J   & I_2 \\
I_2 & I_2 & 0 \\
\end{array}
\right],
}
\eqsa{
\mathcal{J}_{Z_{\rm 2nd}} (\lambda) =
\lambda
\left[
\begin{array}{ccc}
J^{\rm (2nd)}   & I_1^{\rm (2nd)} & I_2^{\rm (2nd)} \\
I_1^{\rm (2nd)} & J^{\rm (2nd)}   & I_2^{\rm (2nd)} \\
I_2^{\rm (2nd)} & I_2^{\rm (2nd)} & K^{\rm (2nd)} \\
\end{array}
\right],
}
\eqsa{
\mathcal{J}_{C_{\rm 3rd}} (\lambda) =
\lambda
\left[
\begin{array}{ccc}
J^{\rm (3rd)} & 0 & 0 \\
0 & J^{\rm (3rd)} & 0 \\
0 & 0 & J^{\rm (3rd)} \\
\end{array}
\right],
}
respectively. The details of the bond are illustrated in Fig.~\ref{fig_Lattice} (a).
Here a parameter $\lambda$ is introduced to
interpolate the {\it ab initio} Hamiltonian for Na$_2$IrO$_3$ at $\lambda=1$
and the Kitaev model at $\lambda=0$.
The {\it ab initio} estimates of the exchange couplings are summarized in Table \ref{abinitioEx}.
This model at $\lambda=1$ well explains not only the thermal properties,
such as the specific heat for $5{\rm K} <T<40{\rm K} $ and the susceptibility for $5{\rm K}<T<400{\rm K}$,
but also the low-lying magnetic excitations~\cite{YYamaji2014,TSuzuki1_2015}.
\begin{table}[ht]
\caption{Exchange couplings of the {\it ab initio} effective Hamiltonian
for Na$_2$IrO$_3$ derived in Ref.~\onlinecite{YYamaji2014}.}
\begin{ruledtabular}
\begin{tabular}{ccccccc}
$\mathcal{J}_Z$ (meV) & $K$ & $J$ & $I_1$ & $I_2$ &  &  \\
                      & -30.7 & 4.4 & -0.4 & 1.1 &  & \\
\hline
$\mathcal{J}_{X,Y}$ (meV) & $K'$ & $J'$ & $J''$ & $I'_1$ & $I'_2$ & $I''_2$  \\
                      & -23.9 & 2.0 & 3.2 & 1.8 & -8.4  & -3.1 \\
\hline
$\mathcal{J}_{Z_{\rm 2nd}}$ (meV) & $K^{\rm (2nd)}$ & $J^{\rm (2nd)}$ & $I_1^{\rm (2nd)}$ & $I_2^{\rm (2nd)}$ &  &  \\
                      & -1.2 & -0.8 & 1.0 & -1.4 &  & \\
\hline
$\mathcal{J}_{C_{\rm 3rd}}$ (meV) & $J^{\rm (3rd)}$ &  &  &  &  &  \\
                      &  1.7 &  &  &  &  & \\
\end{tabular}
\end{ruledtabular}
\label{abinitioEx}
\end{table}

While the ground state of the interpolated Hamiltonian 
is \tm{in} the gapless QSL phase at $\lambda=0$,
the zigzag-type antiferromagnetic order
is stabilized at $\lambda=1$~\cite{YYamaji2014}.
Then, a quantum phase transition from the topological QSL state to the magnetic ordered state
\tm{has to occur at least once, between $\lambda =0$ and 1.}

\subsection{Specific heat}
\label{subsection:SHC}
The specific heat of the spin Hamiltonians, $\hat{H}_{\rm CJK}$ and $\hat{H}_{\lambda}$,   
is calculated by using exact energy spectra up to $N=16$ sites, 
and is estimated by employing thermal pure quantum states~\cite{SSugiura2012,SSugiura2013}
for the 24- and 32-site clusters with the periodic boundary condition.
The finite size clusters used in the following are illustrated in Fig.\ref{fig_Lattice}(b)-(e).

Here we briefly summarize the construction of thermal pure quantum (TPQ) states following Ref.~\onlinecite{SSugiura2012}.
A TPQ state at infinite temperatures is simply given by a random vector, 
\begin{eqnarray}
\ket{\phi_{+\infty}}=\sum_{i=0}^{2^N-1}c_i\ket{i},
\end{eqnarray}
where $\ket{i}$ is \tm{represented by the real-space $S=1/2$ basis and specified by a binary representation of decimal}
and $\{c_i\}$ is a set of random complex numbers with the normalization condition $\sum_{i=0}^{2^N-1} |c_i|^2=1$. 
Then, by utilizing the Lanczos steps with a Hamiltonian $\hat{H}$, the TPQ states at lower temperatures are
constructed as follows:
Starting with an initial vector $\ket{\Phi_0}=\ket{\phi_{+\infty}}$, the $k$-th step Lanczos vector $\ket{\Phi_k}$ $(k\geq 1)$ is constructed as
\begin{eqnarray}
\ket{\Phi_k}=\frac{\hat{H}\ket{\Phi_{k-1}}}{\sqrt{\bra{\Phi_{k-1}}\hat{H}^2\ket{\Phi_{k-1}}}}.
\end{eqnarray}
\tr{The above $k$-th step Lanczos vector is a TPQ state at a finite temperature $T$. 
The corresponding inverse temperature \cyan{${\beta}=(k_{B}T)^{-1}$} is determined through the following formula~\cite{SSugiura2012},}
\begin{eqnarray}
\beta=\frac{2k_{B} k}{\Lambda-\bra{\Phi_{k}}\hat{H}\ket{\Phi_{k}}}+O(1/N),
\end{eqnarray}
where $k_{B}$ is the Boltzmann constant and $\Lambda$ is a constant larger than maxima of $\langle \hat{H} \rangle$.
In other word, a TPQ state at $T$ is given as,
\begin{eqnarray}
\ket{\phi_{T}}=\ket{\Phi_k}.
\end{eqnarray}

The specific heat and entropy of $\hat{H}$ are then estimated by using TPQ states $\ket{\phi_{T}}$.
The thermodynamics and statistical mechanics
\tm{tell us several prescriptions} to calculate the specific heat and the entropy.
Here, we calculate the specific heat $C$ by using the derivative of internal energy with respect to the temperature as
\begin{eqnarray}
C=\frac{d \bra{\phi_{T}}\hat{H}\ket{\phi_{T}}}{d T},\label{def_C}
\end{eqnarray}
%
%
which \tm{is empirically known to be intruded by less} statistical errors in comparison with results obtained through thermal fluctuations of $\hat{H}$.
In the present paper, the entropy $S$ is estimated by integrating $C/T$ from high temperatures as
\begin{eqnarray}
S=Nk_{B}\ln2-\int_{T}^{+\infty}dT' \frac{C}{T'},\label{def_S}
\end{eqnarray}
where
$C\propto T^{-2}$ is assumed in the above integral for the high temperature asymptotic behavior of $C$.
Here, we note that,
for the specific heat and entropy defined in Eq.(\ref{def_C}) and Eq.(\ref{def_S}),
respectively, of the lattice models, it is convenient to use $Nk_{B}$, instead of the gas constant $R$
used in experiments.

\subsection{Equal-time spin correlation}


In comparison with the peak structures of the specific heat,
we examine temperature dependence of the equal-time spin correlations.
For short-range spin correlations, we calculate expectation values of
spin operators $\hat{S}^{\mu}_i \hat{S}^{\mu}_j$ at a finite temperature $T$ 
with the thermal pure quantum states~\cite{SSugiura2012,SSugiura2013}
$\left| \phi_{T} \right\rangle$ as
\begin{eqnarray}
\left\langle \hat{S}^{\mu}_i \hat{S}^{\mu}_j \right\rangle_{T} \equiv
\left\langle \phi_{T} \right| \hat{S}^{\mu}_i \hat{S}^{\mu}_j \left| \phi_{T} \right\rangle,
\end{eqnarray} 
for the nearest-neighbor pairs $\langle i,j\rangle$.
Long-range spin correlations are characterized by the peak \tm{value}
\tm{in the momentum dependence of the equal-time spin structure factor  $S_{T}({\boldsymbol q})$ defined by}
Fourier transformation of $\left\langle \hat{S}^{\mu}_i \hat{S}^{\mu}_j \right\rangle_{T}$,
as
\begin{eqnarray}
S_{T}({\boldsymbol q})=
\frac{1}{N}\sum_{\mu=x,y,z}\sum_{\ell=0}^{N-1}\left\langle \hat{S}^{\mu}_0 \hat{S}^{\mu}_{\ell}\right\rangle_{T} \cos ({\boldsymbol q}\cdot {\boldsymbol R}_{\ell}),
\end{eqnarray}
at $\boldsymbol q=Q$ where ${\boldsymbol R}_{\ell}$ is the position vector of the $\ell$-th site \tm{and $\boldsymbol Q$ is the momentum at the maximum}.

\subsection{Dynamical spin structure factors}
\label{subsection:ME}


To discuss magnetic excitations by a spin flip, we focus on the dynamical spin structure factor (DSF) at zero temperature.
The DSF is defined as
%
\tm{
\begin{eqnarray}
S^{\mu \nu}({\boldsymbol Q},\omega) \equiv -\frac{1}{\pi} \lim_{\epsilon \rightarrow +0} {\rm Im} \langle \phi_0 |
\hat{S}^{\mu \dagger}_{\boldsymbol Q} \frac{1}{\omega+E_0+i\epsilon-{\mathcal H}} \hat{S}^{\nu}_{\boldsymbol Q} |\phi_0\rangle,\nonumber\\
\label{DSF}
\end{eqnarray}%
}
where $\phi_0$ is the ground state of ${\mathcal H}$ with the ground state energy $E_0$. 
The spin operator $\hat{S}^{\mu}_{{\boldsymbol Q}}$ is the Fourier transform of $\hat{S}^{\mu}_{i}$, where $\mu,\nu$ stand for $x,y$ or $z$ component.
After calculating $\phi_0$ and $E_0$
by the Lanczos method, $S^{\mu \nu}({\boldsymbol Q},\omega)$ is obtained by the continued fraction expansion~\cite{EGagliano1987,EDagotto1994}. 

In this paper, we focus on the sum of diagonal elements, namely $\displaystyle S({\boldsymbol Q},\omega)=\sum_{\mu=x,y,z}S^{\mu \mu}({\boldsymbol Q},\omega)$.
In the generalized Kitaev-Heisenberg model, the symmetry of the model Hamiltonian ensures that the off-diagonal component of the DSF becomes exactly zero.
However, in general, $S^{\mu \nu}({\boldsymbol Q},\omega)$ may have non-zero off-diagonal elements,
 if the off-diagonal elements of $\hat{\mathcal J}^{\mu \nu}_{\Gamma_p}$ are non-zero.
The contribution from such off-diagonal element
\tm{is} proportional to the \tm{Fourier transform of the corresponding time-displaced spin correlation}, 
such as $\left\langle \hat{S}^{x}_i(t) \hat{S}^{y}_j(0) \right\rangle$ and $\left\langle \hat{S}^{x}_i(t) \hat{S}^{z}_j(0) \right\rangle$. 
\tm{Since the amplitude of the spin correlation is scaled by the amplitude}
of the matrix element of $\hat{\mathcal J}^{\mu \nu}_{\Gamma_p}$,
\tm{the diagonal elements are dominant.}
\tm{The diagonal elements of the {\it ab initio} Hamiltonian is indeed dominant over the off-diagonal elements, 
and $S({\boldsymbol Q},\omega)$ is expected to contain the main contribution of the spin excitations.}

%
%
%
\section{Thermal and spin excitations}
\label{ResultsDiscussion}
\subsection{Results of generalized Kitaev-Heisenberg model}
\label{subsection:RgKH}

%
\begin{figure*}[htb]
  \begin{center}
   \includegraphics[width=0.8\linewidth]{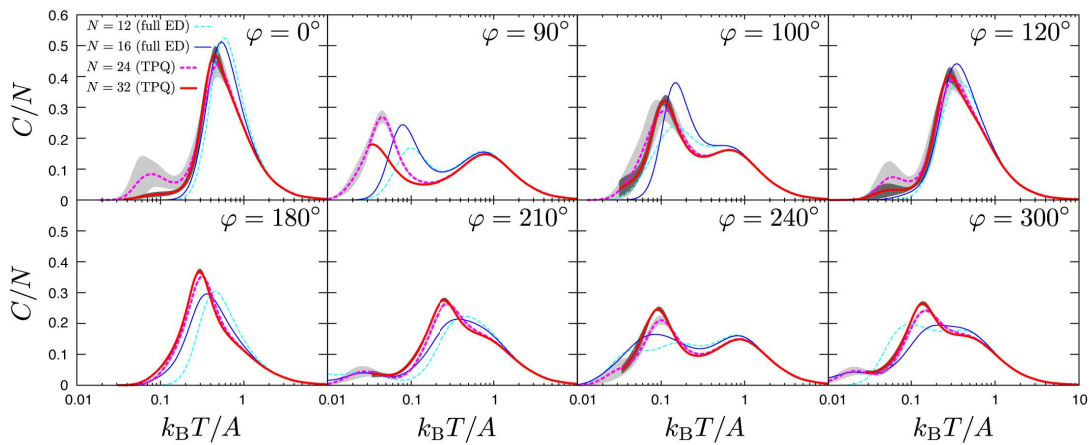}
  \vspace{0.5cm}
\caption{\label{fig_TdepC_gKH} (Color online)
Temperature dependences of the specific heat $C$
of the generalized Kitaev-Heisenberg model $\hat{H}_{\rm CJK}$.
The results for $N=12$ and $N=16$ obtained by fully diagonalizing $\hat{H}_{\rm CJK}$
(denoted as ``full ED") are illustrated with \cyan{thin} broken and \cyan{thin} solid (blue) curves, respectively.
For $N=24$ and $N=32$ (\cyan{thick} broken and \cyan{thick} solid (red) curves),
the thermal pure quantum (TPQ) states~\cite{SSugiura2012} are employed.
The possible errors of TPQ due to the truncation of the Hilbert space are shown in $C$
by shaded (gray) belts, which
is estimated by using the standard deviation of the results obtained from 4 to 36
initial random wave functions at
the high temperature limit, $k_{B}T/A \rightarrow + \infty$.
From the top leftmost to top rightmost panels, $C/N$
is shown for $\varphi=0^{\circ}$,
$90^{\circ}$, $100^{\circ}$, and $120^{\circ}$ in this order.
The same quantities are shown
for $\varphi=180^{\circ}$,
$210^{\circ}$, $240^{\circ}$, and $300^{\circ}$ in this order from the bottom leftmost to
bottom rightmost.}
  \end{center}
\end{figure*}
\begin{figure}[htb]
  \begin{center}
   \includegraphics[width=1\linewidth]{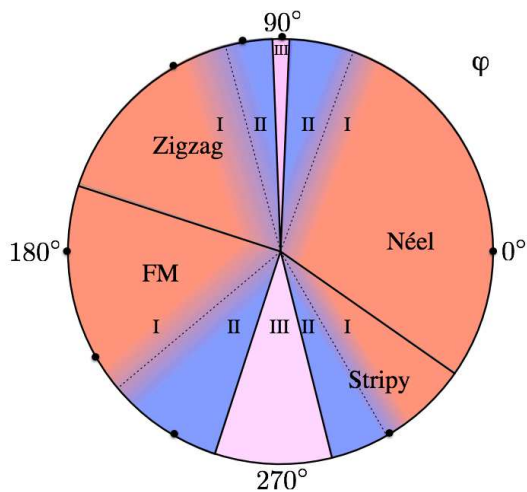}
\caption{\label{fig_phasediagram_gKH} (Color online) Categorization of ground states of
generalized Kitaev-Heisenberg model. 
\tm{Phase boundaries depicted by the solid lines are}  drawn by using the results in Ref. ~\onlinecite{JChaloupka2013}. 
\tm{Solid dots represent} the parameters shown in Figs. \ref{fig_TdepC_gKH}, \ref{fig_TdepSQ_gKH}, and \ref{fig_SQW_gKH}.
Dashed lines
separate whether or not the two-peak structure \tm{in the temperature dependence of the specific heat} is observed within the magnetic ordered phase, \tm{namely the crossover border between the categories I (blue (\textcolor{black}{lightly} shaded) area) and
II (red (\textcolor{black}{dark} shaded) area)}.
\tm{Note that the dotted lines} do {\it not} represent the phase boundary.} 
  \end{center}
\end{figure}
\begin{figure*}[hbt]
  \begin{center}
  \vspace{0.5cm}
   \includegraphics[width=0.8\linewidth]{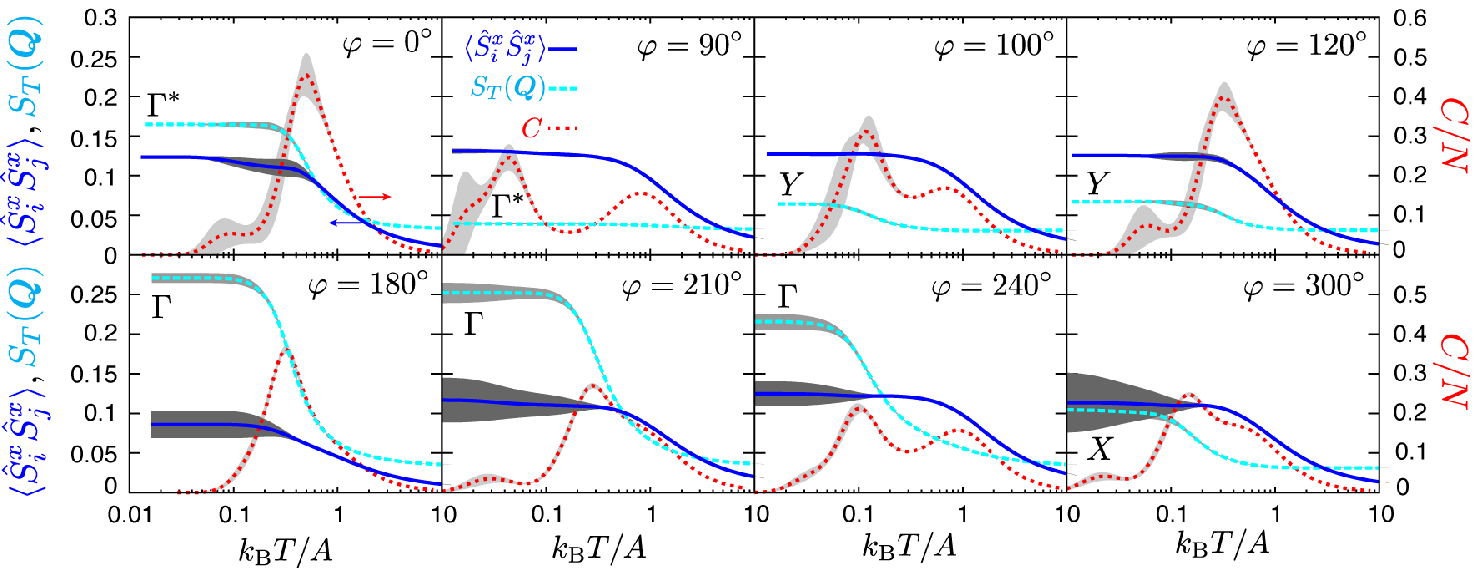}
\caption{\label{fig_TdepSQ_gKH} (Color online)
Temperature dependence of short-range spin correlation $\left\langle \hat{S}^{x}_i \hat{S}^{x}_j \right\rangle_{T}$ for $X$-bond
and long-range spin correlation $S_T({\boldsymbol Q})$ in comparison with the specific heat $C$.
The results for $N=24$ by employing the thermal pure quantum states~\cite{SSugiura2012}
are shown.
The possible errors of TPQ due to the truncation of the Hilbert space are shown in
$\left\langle \hat{S}^{x}_i \hat{S}^{x}_j \right\rangle_{T}$,
$S_T({\boldsymbol Q})$, and $C$
by shaded (gray) belts, which
is estimated by using the standard deviation of the results obtained from 4 to 36
initial random wave functions at
the high temperature limit, $k_{B}T/A \rightarrow + \infty$.
For $S_{T}({\boldsymbol Q})$, following points in Fig.~\ref{fig_SQW_gKH}(i) are selected as the wave vector ${\boldsymbol Q}$;
the $\Gamma$, $Y$, $\Gamma^{\ast}$, and $X$ points are selected \cyan{since $S_{T}({\boldsymbol Q})$ for each value of $\varphi$ has maxima at the momentum}.
These points \tm{represent the Bragg points of}
the ferromagnetic (FM), zigzag, N${\rm \acute{e}}$el, and stripy orders, respectively. \tm{The momentum ${\boldsymbol Q}$ is consistent with the phase diagram in Fig.\ref{fig_phasediagram_gKH}.}
From the top leftmost to top rightmost panels, the results are shown for $\varphi=0^{\circ}$,
$90^{\circ}$, $100^{\circ}$, and $120^{\circ}$ in this order.
The same quantities are shown for $\varphi=180^{\circ}$, $210^{\circ}$, $240^{\circ}$, and $300^{\circ}$
 in this order from the bottom leftmost to bottom rightmost.}
  \end{center}
\end{figure*}

To gain insights into the nature and proximity of the Kitaev's QSL, we compare the specific heat,
the linear spin wave dispersion, and the dynamical spin structure factor (DSF),
\tm{which allows a classification} of the spin Hamiltonians $\hat{H}_{\rm CJK}$ to three distinct categories as summarized in Table \ref{table:category}.
For several choices of the parameter $\varphi$ for the generalized Kitaev-Heisenberg model, 
we classify the nature of the spin and thermal excitations.
The resultant categorization is summarized in Table \ref{table:cgKH}.

The specific heat $C$ for the generalized Kitaev-Heisenberg model is shown in Fig.~\ref{fig_TdepC_gKH}.
Here, we show the results for $N=12$, 16, 24, and 32.
Except the Kitaev's QSLs at $\varphi=90^{\circ}$,
we see no strong $N$ dependence for \cyan{the second-largest and largest} system sizes $N = 24$ and 32. 
For the trivially ordered states at $\varphi=0^{\circ}$ and $180^{\circ}$,
the temperature dependences of $C$ shows a Schottky-like single peak within the error bars.
In the thermodynamic limit, the peak may evolve
into the anomaly (divergence or peak) expected by
the growth of spin correlations accompanied by the transition to the long-range order.
In contrast, for the Kitaev's QSL states at $\varphi=90^{\circ}$ ( and equivalently at $\varphi=270^{\circ}$), 
there are two peaks in the temperature dependences of $C$,
which is a hallmark of thermal fractionalization proposed in Ref.~\onlinecite{JNasu2015}, as is introduced in Sec. \ref{sec:Introduction}.
The low-temperature peak of $C$ may be associated with the contribution from the thermal flux excitations (or the thermal excitations of the localized Majorana fermions),
\tm{while the high-temperature peak may represent the excitation of the itinerant Majorana fermions}.
The category II represented by \cyan{$\varphi=100^{\circ}$ and $240^{\circ}$}
is the same as the category I as to the presence of the magnetic order,
while the two-peak structure of $C$ exists similarly to the category III.
\cyan{The ordered state at $\varphi=300^{\circ}$ is located almost on the border between the category I and category II.}
Although the two-peak structure itself is not a unique feature of the Kitaev's QSL, 
we will propose later that the entropy at temperatures between the two peaks in $C$ serves as a hallmark of the closeness to the Kitaev's QSL.
In Fig.~\ref{fig_phasediagram_gKH}, we show the schematic \tm{illustration for the categorization} obtained from the temperature dependence of $C$ and the presence of the magnetic order.

In Fig.~\ref{fig_TdepSQ_gKH}, we compare the temperature dependence of
the long-range part of the spin correlation represented by
the peak in $S_T({\boldsymbol Q})$ and the short-range part represented by
$\left\langle \hat{S}^{\mu}_i \hat{S}^{\mu}_j \right\rangle_{T}$.
While, in the category I,
the short-range spin correlations $\left\langle \hat{S}^{\mu}_i \hat{S}^{\mu}_j \right\rangle_{T}$ and
the long-range spin correlations $S_{T}({\boldsymbol Q})$ at the ordering wave vector ${\boldsymbol Q}$ grow simultaneously
around the temperature where $C$ has the single peak,
$\left\langle \hat{S}^{\mu}_i \hat{S}^{\mu}_j \right\rangle_{T}$ and $S_{T}({\boldsymbol Q})$ grow independently in the category II represented by
\cyan{$\varphi=100^{\circ}$ and $240^{\circ}$.}
\cyan{The growth of the spin correlation changes from the category I to the category II around $\varphi=300^{\circ}$.}
In the category II, $\left\langle \hat{S}^{\mu}_i \hat{S}^{\mu}_j \right\rangle_{T}$ grows as temperature falls to $T_{h}$, which corresponds to the high-temperature peak in $C$, and saturates below $T_{h}$.
On the other hand,
the long-range spin correlations represented by $S_{T}({\boldsymbol Q})$ grow significantly 
around the temperatures $T_{\ell}$ where $C$ has the low-temperature peak, in the category II.
In the category III, the short-range spin correlation grows at $T_{h}$,
while the long-range part does not show appreciable temperature dependence even at $T_{\ell}$ in contrast to the category  II.
The low temperature peak in $C$ in the category III arises from
an entirely different mechanism from that in the category II, as we already mentioned.
The difference between the category II and the category III is evident in
the temperature dependence of \cyan{the peak of} $S_{T}({\boldsymbol Q})$ in comparison with that of $\left\langle \hat{S}^{\mu}_i \hat{S}^{\mu}_j \right\rangle_{T}$,
as shown  in Fig.\ref{fig_TdepSQ_gKH}.

Based on the above results,
we categorize the quantum phases obtained for the generalized Kitaev-Heisenberg model.
The summary is shown in Table \ref{table:cgKH} and Fig.~\ref{fig_phasediagram_gKH}.
\begin{table}[htb]
\caption{
Categorization of ground states for several choices of $\varphi$
of the generalized Kitaev-Heisenberg model.
Definition of the categories I -- III is shown in Table \ref{table:category}.}
\begin{tabular}{clc}
\hline
\hline
$\varphi$ & quantum phase & category \\
\hline
$0^{\circ}$ & N${\rm \acute{e}}$el & I. \\ 
$90^{\circ}$ & Kitaev's QSL & III. \\ 
$100^{\circ}$ & zigzag & II. \\ 
$120^{\circ}$ & zigzag & I. \\ 
$180^{\circ}$ & ferromagnetic & I. \\ 
$210^{\circ}$ & ferromagnetic & I. \\ 
$240^{\circ}$ & ferromagnetic & II. \\ 
$270^{\circ}$ & Kitaev's QSL & III. \\ 
$300^{\circ}$ & stripy & \cyan{I/II.} \\ 
\hline
\hline
\end{tabular}
\label{table:cgKH}
\end{table}

\begin{figure*}[htb]
  \begin{center}
   \includegraphics[width=0.8\linewidth]{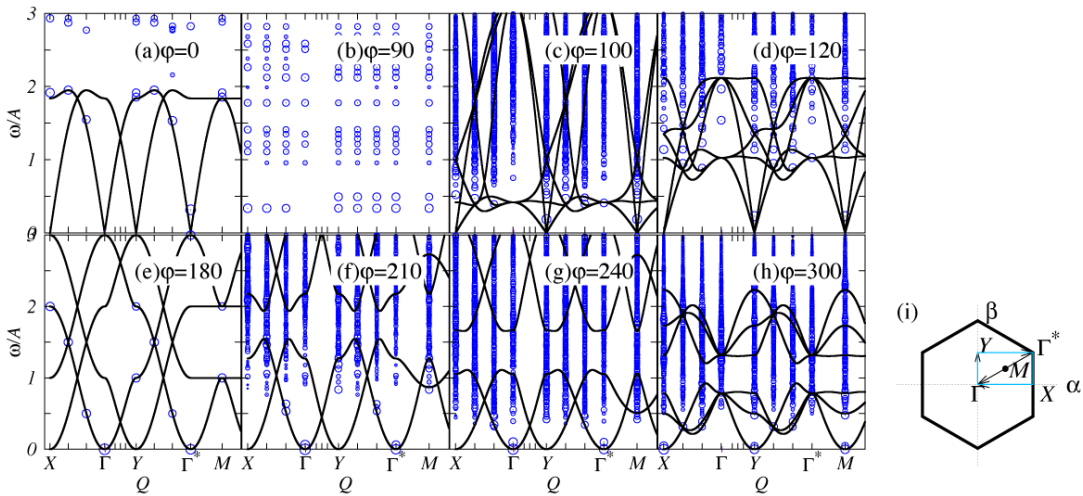}
     \vspace{0.5cm}
\caption{\label{fig_SQW_gKH} (Color online) Dynamical spin structure factors of
generalized Kitaev-Heisenberg models for 24-site cluster.
Area of each circle is proportional to the intensity in logarithmic scale.
Abscissa represents labeled points in (i).
From the top leftmost to top rightmost panels,
$S({\boldsymbol Q},\omega)$ is shown for (a) $\varphi=0^{\circ}$,
(b) $90^{\circ}$, (c) $100^{\circ}$, and (d) $120^{\circ}$ in this order.
$S({\boldsymbol Q},\omega)$ is shown for (e) $\varphi=180^{\circ}$,
(f) $210^{\circ}$, (g) $240^{\circ}$, and (h) $300^{\circ}$ in the order from
the bottom leftmost to bottom rightmost.
Here, the results of the linear spin wave theory are shown in solid curves.
In order to compare the spin wave dispersions with the low-lying excitations, each spin wave result in the top panels is multiplied by a constant.
The multiplication constants for (a), (c) and (d) are 1.3, 1.65, and 1.45, respectively. The spin wave results in the lower panels are drawn without such tuning constant.
For the results in the zigzag and stripy phase, (c), (d), and (h), we also plot the spin wave dispersions obtained from the case where the ordered state is rotated by $2\pi/3$.}
  \end{center}
\end{figure*}

By comparing the low-lying excitations of  the dynamical spin structure factors (DSFs) with the linear spin wave approximation, 
we further confirm \tm{that the categorization is robust}.
In Fig.\ref{fig_SQW_gKH}, we show both results of the DSF $S({\boldsymbol Q},\omega)$ for $N=24$ and of the linear spin wave calculations.

We start from the results for the simplest case.
At $\varphi=0^{\circ}$ and $180^{\circ}$, the model becomes the antiferromagnetic/ferromagnetic Heisenberg model.
Therefore, the low-lying excitation of the DSF is expected to be well explained by the spin wave mode.
At $\varphi=180^{\circ}$, all poles in the DSF \tm{are perfectly located} on the spin wave mode.
In the antiferromagnetic case at $\varphi=0^{\circ}$,
the low-lying excitation agrees with the linear spin wave mode by introducing a renormalization factor $a$.
$a$ is estimated from the best fitting of spin wave dispersions for the poles of the low-lying excitations in the DSF; 
$S({\boldsymbol Q},\omega_{\rm lowest}) \sim a \times \omega_{\rm LSW}({\boldsymbol Q})$,
where $S({\boldsymbol Q},\omega_{\rm lowest})$ denotes the poles of the lowest excitation in the DSF and $\omega_{\rm  LSW}({\boldsymbol Q})$ is the linear spin wave mode.
At $\varphi=0^{\circ}$, we obtain $a \sim 1.3$, which is the upper limit because the positions of poles are affected by the system size.
(The size dependence becomes strong especially at the symmetric wave-number points. )
As well studied in Refs.~[\onlinecite{PhysRev.128.2131,PhysRevB.39.9760}], 
the exact low-lying excitations in the Heisenberg models are described by the linear spin wave mode with an $O(1)$ renormalization factor $a$. 
The renormalization factor is $a = \frac{\pi}{2}$ in the S=1/2 spin chain case~\cite{PhysRev.128.2131}. 
This value can be an indicator for the renormalization of the quantum fluctuation. 
As shown below, we observe the two-peak structure in the temperature dependence of the specific heat $C$, when $a \gtrapprox 1.5$.

In contrast, magnetic excitations described by the poles in the DSF are
completely different from coherent magnetic excitations in the Kitaev's QSL phase at $\varphi=90^{\circ}$($270^{\circ}$).
This is due to diverging quantum fluctuations in the Kitaev's QSL phase and
 massive degeneracy of classical spin orders spoils the linear spin wave analysis.
Below, we explain the low-lying excitations of the DSFs, when the system approaches
the Kitaev's QSL phase from the deep inside of the magnetic ordered phase.

First, we focus on the positive Kitaev-coupling case for $0\le \varphi \le 180^{\circ}$.
For  $0\le \varphi \le 180^{\circ}$ in the magnetic ordered phases, the low-lying excitation of the DSF can be explained by the correlated/renormalized magnon excitation. 
The fitting parameter $a$ is always larger than the unity; the low-lying excitation becomes hard in comparison with the linear spin wave mode. 
When the system approaches
the Kitaev's QSL phase around $90^{\circ}$, $a$ drastically increases.
\tm{Though} not shown in \tm{the} figure, we obtain
\tm{$a \sim 1.85$}  at $\varphi=86^{\circ}$.
At $\varphi=100^{\circ}$ in the category II,  the ground state is the zigzag ordered state. 
From Fig. \ref{fig_TdepC_gKH},
we confirm the two-peak structure in the temperature dependence of $C$.
The renormalization factor $a$ at $\varphi=100^{\circ}$ is estimated as $a \sim 1.65$ and is larger than that of the S=1/2 chain case.
When the system goes into the deep inside of the magnetic phase, 
 the factor $a$ decreases and crosses $a \sim 1.5$, where the two-peak structure is almost  smeared out in $C$.
At $\varphi=120^{\circ}$ in the category I, $a$ is about 1.45.
While the ground state is still in the zigzag ordered phase, 
the temperature dependence of $C$ shows the usual single peak.

Next, we see the results for the negative Kitaev coupling case for $180^{\circ} \le \varphi \le 360^{\circ}$. 
In contrast to the positive Kitaev coupling case, 
the low-lying excitation in the DSF can be well explained by free magnon picture for
\textcolor{black}{$K<0$}
in the magnetic ordered phases:
\textcolor{black}{T}he spin wave mode except the M/$\Gamma$($\Gamma^*$) point can explain
 the low-lying excitations of the DSF without the renormalization factor $a$ discussed above.

\textcolor{black}{Here, we detail the discrepancy between the spin wave mode and the low-lying excitation of the DSF
for $K<0$, which is another clue to categorize the magnetic ordered phases into the categories I and II
in addition to the temperature dependence of specific heat.}
When the system approaches the Kitaev's QSL phase around $\varphi=270^{\circ}$, 
the categorization of the categories I and II can be discussed from the discrepancy 
between the spin wave mode and the low-lying excitation of the DSF at the M point.
At $\varphi=210^{\circ}$ in the category I, the low-lying excitation is well explained by the spin wave mode.
However, the discrepancy between the low-lying excitation of the DSF and the spin wave mode develops 
clearly at the M point at $\varphi=240^{\circ}$ in the category II.
The low-lying \tm{excitations} of the DSF are located in the lower energy region than that of the spin wave mode.
(The discrepancy between the both is also clear at the Y point. 
However, the excitation of the DSF at the Y point is identical to that at the M point. 
This is due to to the finite size effect; the symmetry breaking is prohibited in the finite size systems.)
\tm{We regard $\varphi=210^{\circ}$ is located in I near in the crossover region of the categories I and II, where both characters are mixed.} 
Based on these observations, here we categorize $\varphi=210^{\circ}$ and  $\varphi=240^{\circ}$ as the category I and category II, respectively.
In the stripy phase for $\varphi > 270^{\circ}$, the wave vector, where the spin wave mode deviates from the low-lying excitations of the DSF, moves to the $\Gamma$ and $\Gamma^*$ points.
At $\varphi=300^{\circ}$, the low-lying excitation of the DSF shows \cyan{slight} softening at the $\Gamma$ and $\Gamma^*$ points in comparison with the spin wave mode. 
\cyan{In addition, the temperature dependence of $C$ shows a single peak with a prominent shoulder, which is
in between the single peak structure of the category I and the two-peak structure of the category II.  
Consequently, the system at $\varphi=300^{\circ}$ is concluded to be located
\tm{in the crossover region} between the category I and the category II.}

\subsection{Results for {\it ab initio} Hamiltonian of Na$_2$IrO$_3$}
\label{subsection:RiH}
%
%
\begin{figure}[hbt]
  \begin{center}
  \vspace{0.5cm}
   \includegraphics[width=0.85\linewidth]{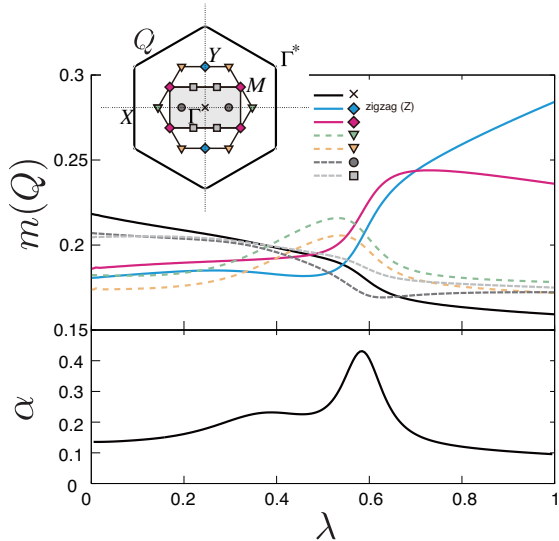}
\caption{\label{figphase} (Color online) Square root of the \tm{equal-time spin} structure factors, $m({\boldsymbol Q})$
and the second derivative of ground state energy $\alpha$ as functions of $\lambda$ calculated with
the $N=24$-site cluster in Fig.\ref{fig_Lattice}(d).}
  \end{center}
\end{figure}
\begin{figure*}[htb]
  \begin{center}
   \includegraphics[width=0.8\linewidth]{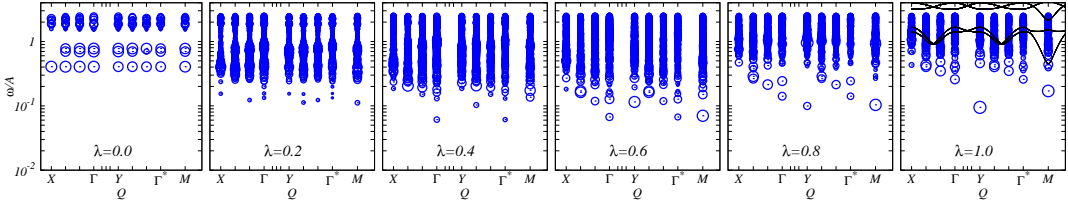}
  \vspace{0.5cm}   
\caption{\label{fig_SQW_ABINITIO} (Color online) Dynamical spin structure factors of
the interpolated Hamiltonian $\hat{H}_{\lambda}$ between the Kitaev limit and
the {\it ab initio} spin Hamiltonian proposed in
Ref.~\onlinecite{YYamaji2014}.
From the leftmost to rightmost panels, $S({\boldsymbol Q},\omega)$ is shown for $\lambda=0$,
$\lambda=0.2$, $\lambda=0.4$, $\lambda=0.6$, $\lambda=0.8$, and $\lambda=1$.
Abscissa represents the labeled points in Fig.\ref{fig_SQW_gKH}(i).}
  \end{center}
\end{figure*}
\begin{figure*}[hbt]
  \begin{center}
  \vspace{0.5cm}
   \includegraphics[width=0.8\linewidth]{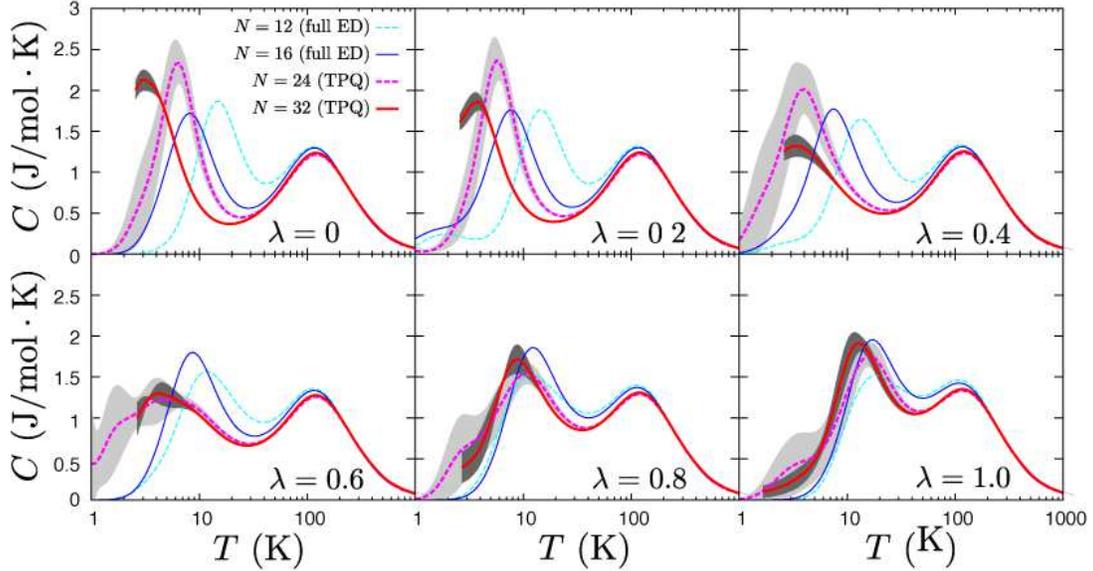}
\caption{\label{fig_TdepC_ABINITIO} (Color online)
Temperature dependences of the specific heat $C$
of the interpolated Hamiltonian $\hat{H}_{\lambda}$ between the Kitaev limit and the
{\it ab initio} spin Hamiltonian proposed in
Ref.~\onlinecite{YYamaji2014}.
The results for $N=12$ and $N=16$ obtained by fully diagonalizing $\hat{H}_{\lambda}$
(denoted as ``full ED") are illustrated with \cyan{thin} broken and \cyan{thin} solid (blue) curves, respectively.
For $N=24$ and $N=32$ (\cyan{thick} broken and \cyan{thick} solid (red) curves), the thermal pure quantum (TPQ) states~\cite{SSugiura2012} are employed.
The possible errors of TPQ due to
the truncation of the Hilbert space are shown in $C$ by shaded (gray) belts,
which is estimated by using the standard deviation of the results obtained from 4 to 36
initial random wave functions.}
  \end{center}
\end{figure*}
\begin{figure*}[hbt]
  \begin{center}
  \vspace{0.5cm}
   \includegraphics[width=0.7\linewidth]{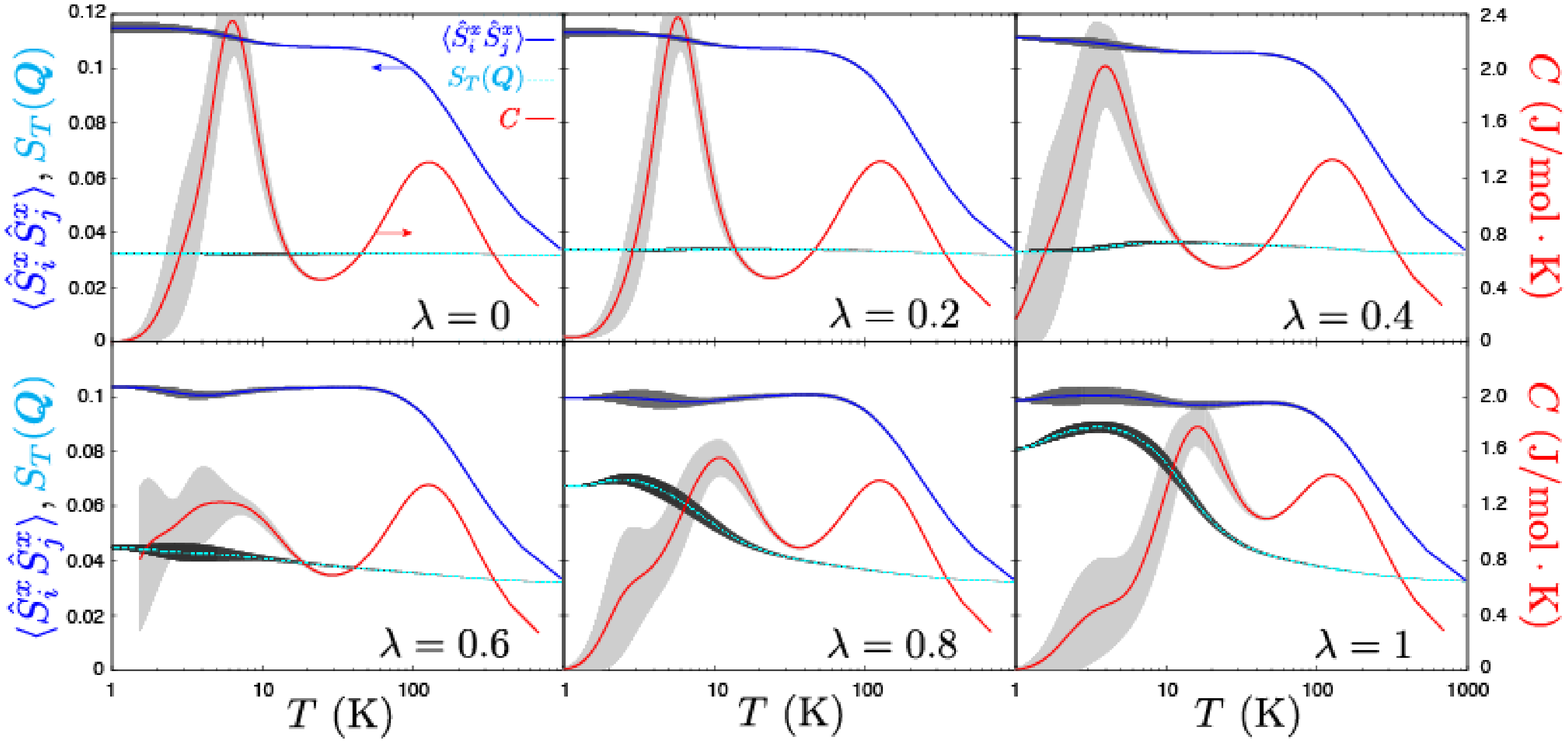}
\caption{\label{fig_abSS24} (Color online)
Temperature dependence of the short-range spin correlation
$\left\langle \hat{S}^{x}_i \hat{S}^{x}_j \right\rangle_{T}$ for $X$-bond
and the long-range spin correlation $S_T({\boldsymbol Q})$ in comparison with the specific heat $C$.
The results for $N=24$ by employing the thermal pure quantum states~\cite{SSugiura2012} are shown.
The possible errors of TPQ due to the truncation of the Hilbert space are shown in
$\left\langle \hat{S}^{x}_i \hat{S}^{x}_j \right\rangle_{T}$,
$S_T({\boldsymbol Q})$, and $C$
by shaded (gray) belts, which
is estimated by using the standard deviation of the results obtained from 4 to 36
initial random wave functions.
For $S_{T}({\boldsymbol Q})$, the Y point is selected as the wave vector ${\boldsymbol Q}$, 
which corresponds to the zigzag (Z) order~\cite{YYamaji2014}. }
  \end{center}
\end{figure*}

In the light of the categorization examined in the generalized Kitaev-Heisenberg models,
we examine the interpolated Hamiltonian between the Kitaev limit, $\hat{H}_{\lambda=0}$,
and the {\it ab initio} Hamiltonian of Na$_2$IrO$_3$, $\hat{H}_{\lambda=1}$, given in Eq.(\ref{ham:abinitio}).
First, we show that the \tm{peak value of the equal-time spin} structure factor for the zigzag order starts \tm{growing} at an onset value $\lambda_c$.
\textcolor{black}{Here $\lambda_c $ is around 0.6.
For all $0\leq \lambda \leq 1$, the dynamical spin structure factor of $\hat{H}_{\lambda}$
is not captured by the linear spin wave theory~\cite{TSuzuki1_2015}.
Next, we show that the temperature dependence of the specific heat for $\hat{H}_{\lambda}$
always has a two-peak structure irrespective of $\lambda$.}


The quantum phases of the interpolated Hamiltonian are examined by
\tm{equal-time spin} structure factors and the second derivatives of the ground state energy defined below;
here we introduce the square root of the normalized \tm{equal-time spin} structure factor,
which is extrapolated to the magnetic order parameter at momentum ${\boldsymbol Q}$ in the thermodynamic limit, defined as
\begin{eqnarray}
m({\boldsymbol Q})
\equiv
\lim_{T\rightarrow +0}\sqrt{S_{T}({\boldsymbol Q})},
\end{eqnarray}
\tm{where we take the $T=0$ limit.} 

The quantum phase transitions are expected to cause divergence or discontinuity (or a sharp peak in finite-size systems)
in the second derivatives of the ground state energy $E$ with respect to $\lambda$,
\begin{eqnarray}
\alpha\equiv -\frac{d^2 E/N}{d\lambda^2}. 
\end{eqnarray}
In Fig.\ref{figphase}, the $\lambda$-dependences of $m(\boldsymbol Q)$ and $\alpha$ for $N=24$ site cluster are given.

From the growth of $m(\boldsymbol Q)$ at the Y point corresponding to the zigzag order observed in the experiments and a peak in $\alpha$ at $\lambda \sim 0.59$,
we conclude that the zigzag order appears for $\lambda \gtrsim 0.6$.
For $\lambda \lessapprox 0.4$, we expect the \tr{Kitaev's} QSL ground states.
Since the phase transition 
around $\lambda\sim 0.6$
appears to be continuous with the reduction of $m$ toward the transition point $\lambda\sim 0.6$, 
the distance from the \tr{Kitaev's} QSL phase may be measured from the ordered moment.

The presence of the phase boundary to the zigzag ordered phase is also confirmed from the DSF results shown in Fig.~\ref{fig_SQW_ABINITIO}.
At $\lambda=0$, we observe a characteristic non-dispersive mode at $\omega/A \sim 0.3$ reflecting the Kitaev's QSL ground state~\cite{JKnolle2014}.
For $\lambda<0.6$, some poles with the weak intensity appear below $\omega/A \sim 0.3$ 
and the peak with the largest intensity is not located in the lowest excitation mode.
These properties of the low-energy excitations are contrast to those in the magnetic ordered phase, where the lowest excitation usually shows the largest intensity.
We also observe the lack of well developed peaks in the \tm{equal-time spin} structure factors shown in Fig.~\ref{figphase}.
Thus, we conclude that the ground state at $\lambda < 0.6$ is still the Kitaev's QSL state.


For $0.6 < \lambda \le 1.0$,  the lowest excitation with the largest intensity appears at the M or Y point 
and the peak at the Y point develops as $\lambda$ increases.
In this region, the \tm{equal-time spin} structure factors at the M and Y point well develops as shown in Fig.~\ref{figphase}.
Therefore, the ground state becomes magnetically ordered for $\lambda > 0.6$ and the presence of the phase boundary is expected for $\lambda \sim 0.6$.

To confirm that the interpolated Hamiltonian $\hat{H}_{\lambda}$ is categorized into the category II
\textcolor{black}{for $\lambda>0.6$}, 
we also examine the temperature dependences of the specific heat $C$.
The results are shown in Fig.~\ref{fig_TdepC_ABINITIO}.
First of all, for the entire parameter range, $0\leq \lambda \leq 1$,
two peaks are seen in the temperature dependences of $C$.
\textcolor{black}{The low-temperature peak} is at $T/|K| \lesssim 0.03$, and the higher (high-$T$) one is at $T/|K| \sim 0.3$,
where $|K|$=30.7 meV \tr{(356 K)} (see Table~\ref{abinitioEx}).
As already discussed in Ref.~\onlinecite{JKnolle2014} and Ref.~\onlinecite{JNasu2015}, 
the origin of the high-$T$ peak is well explained by the growth
of magnetic correlations for the nearest neighbor pairs,
which is determined by the Kitaev couplings, $K\sim K'$.
The low-$T$ peak in the pure Kitaev model corresponds to
the thermal fluctuation of the local $Z_2$ gauge field that is
one of two Majorana fermions yielded via the fractionalization of an original quantum spin. 
The non-monotonic $\lambda$-dependences in the low-$T$ peak correspond to
the quantum phase transition around $\lambda \sim 0.6$
and \tr{possible emergence of the intermediate phase for $0.4 \lesssim \lambda \lesssim 0.6$}.

\section{Distance from the Kitaev spin liquid phase}
\label{section:DdK}
%
%
\begin{figure*}[hbt]
  \begin{center}
  \vspace{0.5cm}
   \includegraphics[width=0.6\linewidth]{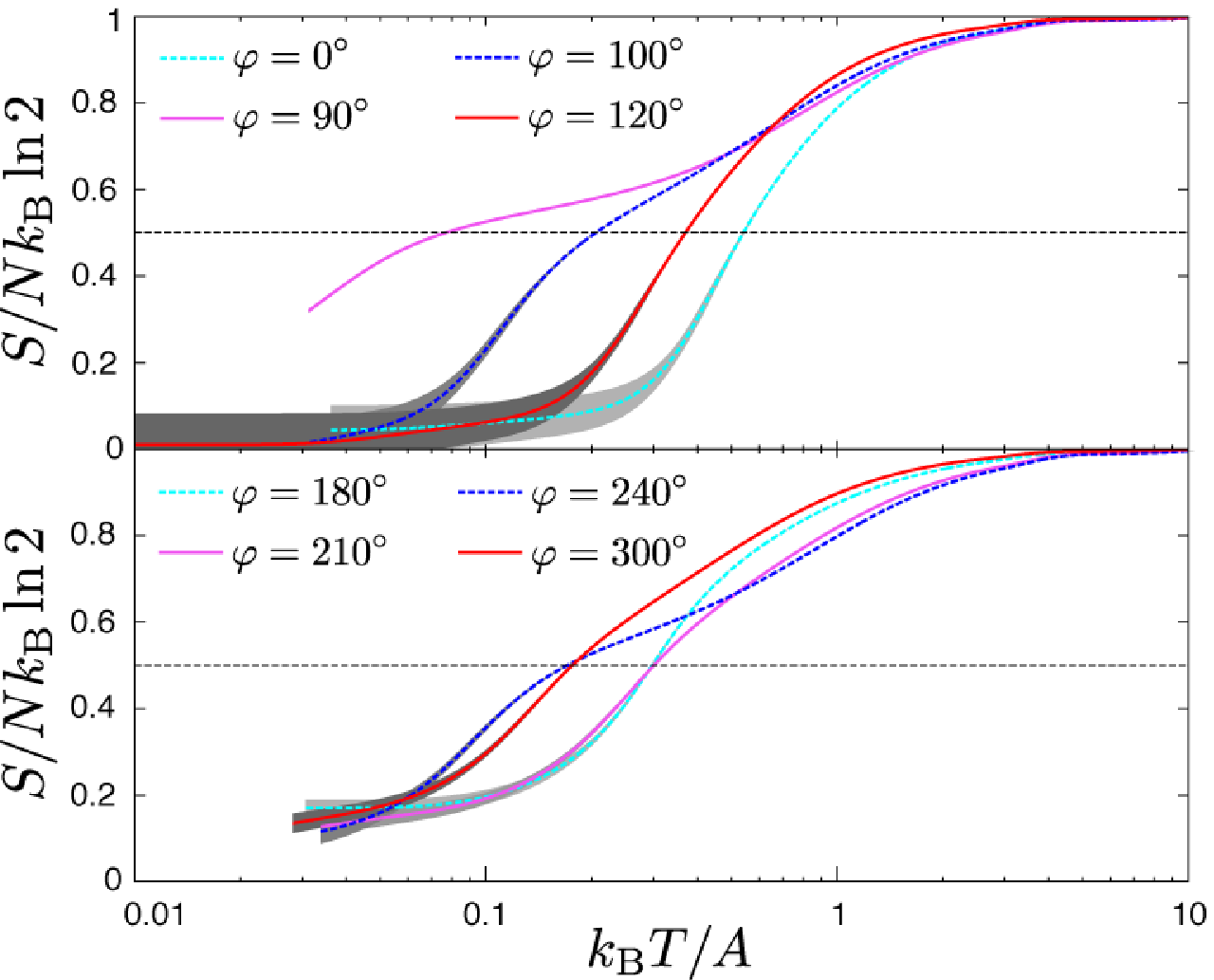}
\caption{\label{fig_TdepS_gKH} (Color online)
Temperature dependence of the entropy for the generalized Kitaev-Heisenberg Hamiltonian.
The horizontal broken line indicates \cyan{$S=(Nk_{B}/2)\ln 2$}.
The possible errors of TPQ due to
the truncation of the Hilbert space are shown in $S$ by shaded (gray) belts,
which is estimated by using the standard deviation of the results obtained from 4
initial random wave functions.
}
  \end{center}
\end{figure*}

\begin{figure}[bth]
  \begin{center}
  \vspace{0.5cm}
   \includegraphics[width=0.8\linewidth]{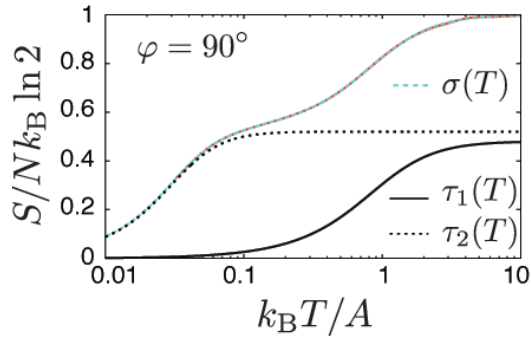}
\caption{\label{fig_decomp_phi} (Color online)
\tbr{Decomposition of $S$ for the generalized Kitaev-Heisenberg Hamiltonian with $\varphi=90^{\circ}$.
The possible errors of TPQ due to
the truncation of the Hilbert space are shown in $S$ by shaded (gray) belts,
which is estimated by using the standard deviation of the results obtained from 4
initial random wave functions.
The function $\sigma (T)(=\tau_1 (T) + \tau_2 (T))$ defined in Eqs.(\ref{fit_sigma}) and (\ref{fit_tau})
is employed to fit the numerical result denoted by the (red) solid curve.
The fitting function $\sigma (T)$ denoted by the broken (light blue) curve is
almost on top of the numerical result of $S$ for $\varphi=90^{\circ}$.
The decomposed components $\tau_1 (T)$ and $\tau_2 (T)$ are shown in
(black) solid and dotted curves, respectively.
}
}
  \end{center}
\end{figure}

\begin{figure}[bth]
  \begin{center}
  \vspace{0.5cm}
   \includegraphics[width=1.0\linewidth]{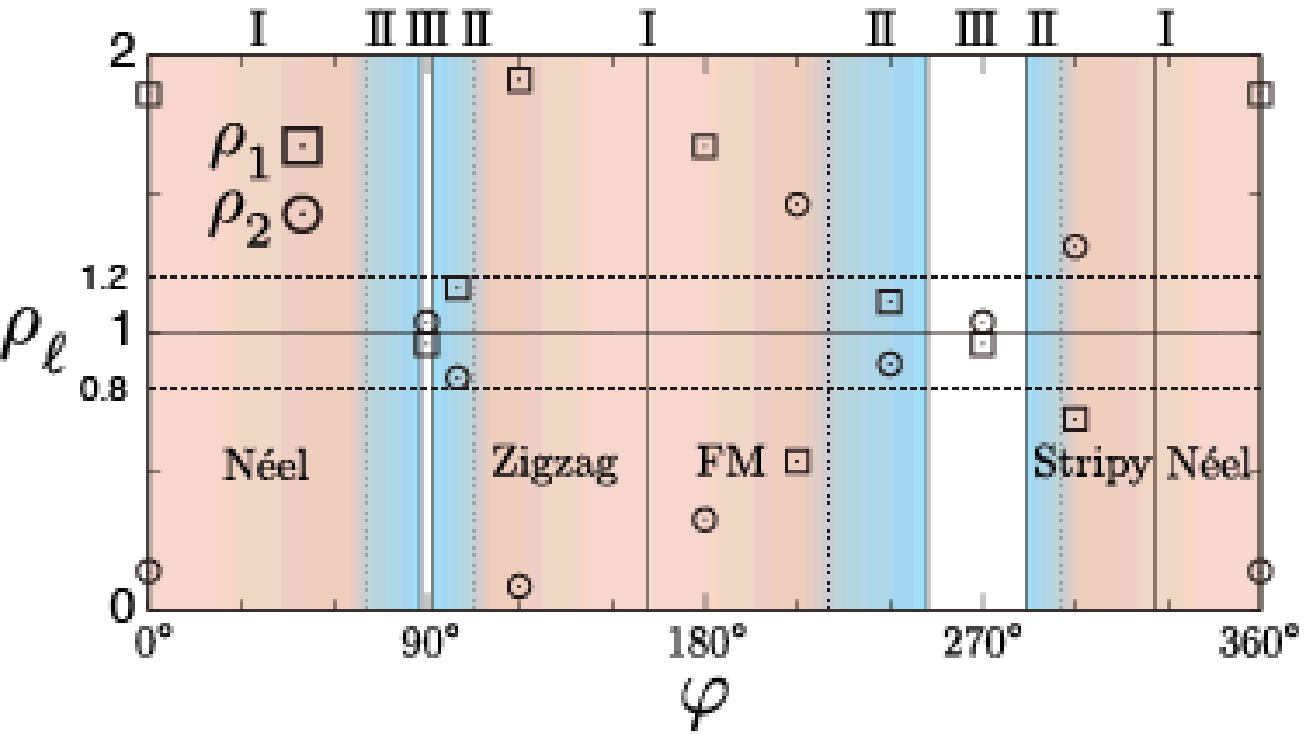}
\caption{\label{fig_rho_phi} (Color online)
Weights $\rho_1$ and $\rho_2$ in least-square fitting of $S(T)/Nk_{B}\ln 2$ with $\sigma (T)$ as
functions of $\varphi$.
}
  \end{center}
\end{figure}

\begin{figure*}[hbt]
  \begin{center}
  \vspace{0.5cm}
   \includegraphics[width=0.65\linewidth]{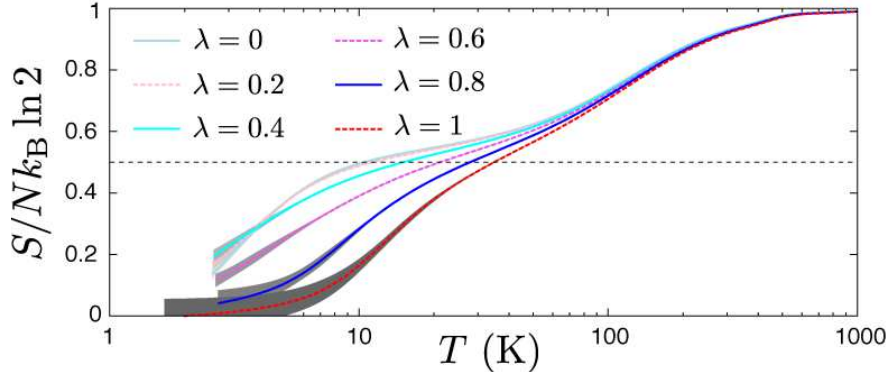}
\caption{\label{fig_S_T_ablambda} (Color online)
Temperature dependence of the entropy for the interpolated Hamiltonian.
$\lambda=0$ ($\lambda=1$) corresponds to the Kitaev limit (the {\it ab initio} Hamiltonian of Na$_2$IrO$_3$).
The results for $N=32$ are shown.
The horizontal broken line indicates \cyan{$S=(Nk_{B}/2)\ln 2$}.
The possible errors of TPQ due to
the truncation of the Hilbert space are shown in $S$ by shaded (gray) belts,
which is estimated by using the standard deviation of the results obtained from 4
initial random wave functions.
}
  \end{center}
\end{figure*}

\begin{figure}[bth]
  \begin{center}
  \vspace{0.5cm}
   \includegraphics[width=0.8\linewidth]{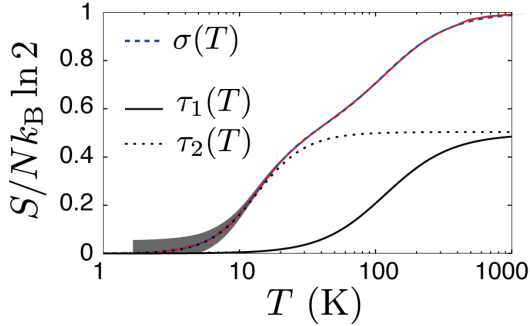}
\caption{\label{fig_decomp_ab} (Color online)
\tbr{Decomposition of $S$ for the interpolated Hamiltonian with $\lambda=1$.
The possible errors of TPQ due to
the truncation of the Hilbert space are shown in $S$ by shaded (gray) belts,
which is estimated by using the standard deviation of the results obtained from 4
initial random wave functions.
The function $\sigma (T)(=\tau_1 (T) + \tau_2 (T))$ defined in Eqs.(\ref{fit_sigma}) and (\ref{fit_tau})
is employed to fit the numerical result denoted by the (red) solid curve.
The fitting function $\sigma (T)$ denoted by the broken (light blue) curve is
almost on top of the numerical result of $S$ for $\varphi=90^{\circ}$.
The decomposed components $\tau_1 (T)$ and $\tau_2 (T)$ are shown in
(black) solid and dotted curves, respectively.
}
}
  \end{center}
\end{figure}
\begin{figure}[bth]
  \begin{center}
  \vspace{0.5cm}
   \includegraphics[width=0.8\linewidth]{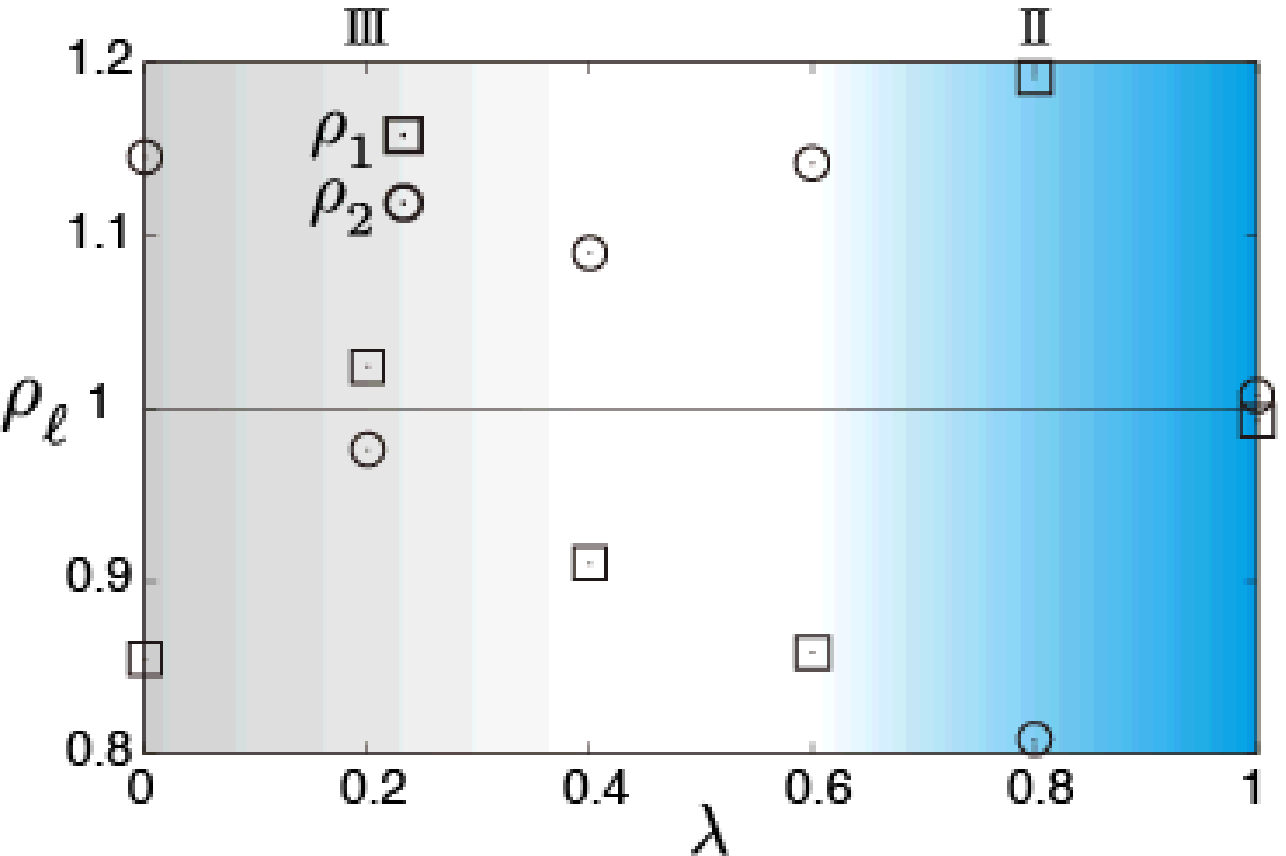}
\caption{\label{fig_rho_ab} (Color online)
Weights $\rho_1$ and $\rho_2$ in least-square fitting of $S(T)/Nk_{B}\ln 2$ with $\sigma (T)$ as
functions of $\lambda$.
}
  \end{center}
\end{figure}
\begin{figure}[bth]
  \begin{center}
  \vspace{0.5cm}
   \includegraphics[width=1.0\linewidth]{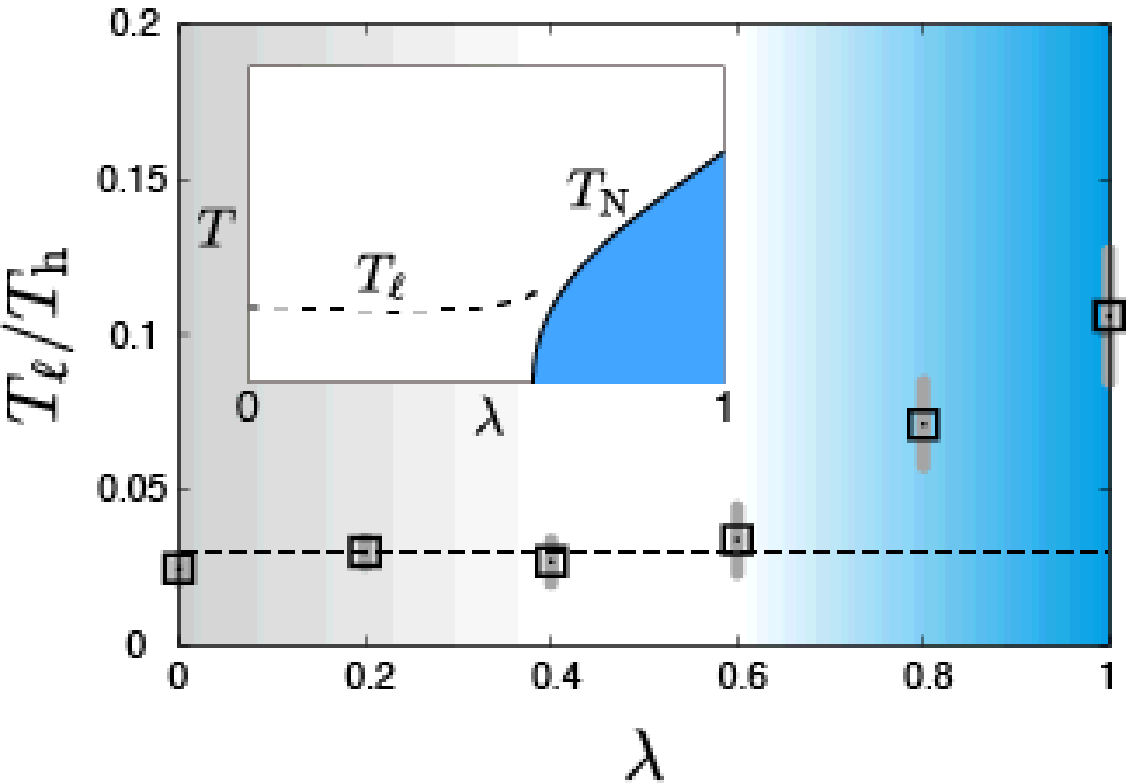}
\caption{\label{fig_phase_diagramv2} (Color online)
Ratio of two temperature scales, \cyan{$T_{\ell}$ and $T_{h}$}.
The ratio \cyan{$T_{\ell}/T_{h}$} $(<1)$ is obtained through the fitting function
$\sigma (T)$ defined in Eqs.(\ref{fit_sigma}) and (\ref{fit_tau}).
The horizontal broken line represents a proposed upper bound for \cyan{$T_{\ell}/T_{h}$}
in the Kitaev's spin liquids.
\cyan{The vertical (grey) bars illustrate uncertainty in $T_{\ell}/T_{h}$ due to
uncertainty in $C(T)$.}
\tr{The shaded areas for $\lambda\gtrsim 0.6$ and $\lambda\lesssim 0.4$ indicate
the zigzag ordered and the Kitaev's QSL ground states, respectively.}
The inset shows schematic phase diagram expected from the present results.
Here, the temperature $T_{\ell}$ corresponds to the lower temperature peak of $C$ in Fig.\ref{fig_TdepC_ABINITIO},
which is denoted as the broken curve.
The $\lambda$-dependence of the expected \tbr{N\'eel} temperature $T_{N}$ for
the zigzag ordered phase is shown in the solid curve.
}
  \end{center}
\end{figure}

In the previous \cyan{section, Sec. \ref{ResultsDiscussion},}
magnetically ordered materials categorized as the category II are expected to be
close to the Kitaev's spin liquid in the parameter space of the effective Hamiltonians.
We further propose more quantitative measure to estimate the distance between the real material and the Kitaev's QSL.

As is shown in Figs.~\ref{fig_TdepC_gKH} and \ref{fig_TdepC_ABINITIO}, 
the specific heat $C$ in the category II shows two peaks similar to the category III.
However, the two-peak structure in the temperature dependence of $C$ itself 
is not a unique feature in the vicinity of the Kitaev's QSL.
For example, geometrically frustrated and quasi-one-dimensional quantum spin systems also show
the two-peak structure in the temperature dependence of $C$~\cite{PhysRevB.68.014424}.
In that case, the high- and low-temperature peaks depend on the dimensional anisotropy,
where the low-temperature peak in $C$ represents the entropy release arising from the real long-range order.
If the anisotropy increases, the release of the entropy at the low temperature peak may decrease.

On the other hand, the two-peak structure of $C$ in the Kitaev's QSL originates
 from the fractionalization of the spin degrees of freedom.
The fractionalization is expected to be more evident in temperature dependence of the entropy $S$ that
directly shows the fractionalized spin degrees of freedom.
A quantitative way of measuring the distance to the Kitaev's QSL is 
to observe the temperature dependences of $S$.
As clarified in Ref.~\onlinecite{JNasu2015},
at temperatures between the two peaks of $C$,
the systems with the Kitaev's QSL ground states show a half plateau in $S$ 
with the value $(Nk_{B}/2)\ln 2$.
The value $(Nk_{B}/2)\ln 2$ is not generically expected in other cases unless some accidental coincidence happens.
Thus, this value is a hallmark of the Kitaev's QSL and persists in the systems in close vicinity of
the Kitaev's QSL phase, even when the systems show magnetically ordered states, as shown in the remaining part.

Before discussing the temperature dependence of the entropy $S$,
we explain the physical mechanism of the two-peak formation in $C$ from the viewpoint of the spin correlation.
The growth of the long-range spin correlations below the low-temperature peak of $C$ clearly distinguishes 
the magnetic ordered ground states in the category II from the Kitaev's QSL in the category III,
as seen in Figs.~\ref{fig_TdepSQ_gKH} and  \ref{fig_abSS24}. 
For example, \tm{in} the generalized Kitaev-Heisenberg Hamiltonian with $\varphi =100^{\circ}$ and $\varphi = 240^{\circ}$
and in the {\it ab initio} Hamiltonian for $\lambda \gtrapprox 0.6$, $S_{T}({\boldsymbol Q})$ \tm{grows with decreasing temperature and} is saturated \tm{only} below the lower temperature peak,
while that  \tm{remains small} in the Kitaev's QSL phase.
On the other hand, the short-range spin correlations
\tm{grows with decreasing temperatures until the saturation} below the high-temperature peak of $C$
for the entire parameter ranges.

We show the temperature dependences of $S$ in Fig.\ref{fig_TdepS_gKH} for the generalized Kitaev-Heisenberg Hamiltonian. 
When the system is categorized as the categories II or III,
 a shoulder around $S=(Nk_B/2)\ln 2$ is observed in the temperature dependence of $S$.
At least, the value $(Nk_B/2)\ln 2$ is due to the contribution from the free Majorana fermions in the category III.
The plateau in $S$ is evident at an anisotropic limit \tr{$|K_z|\gg |K_x| = |K_y|$ $(K\gg K')$}~\cite{JNasu2015},
and is smeared out as the system approaches the symmetric Kitaev couplings, \tr{$|K_z| = |K_x| = |K_y|$ $(K\sim K')$},
which is identical to $\varphi=90^{\circ}$ ($\varphi=270^{\circ}$).
However, the remaining feature, namely the shoulder structure, is still \tm{an} evidence \tm{for} the fractionalization of the quantum spins.
In the present case, we still observe such shoulder structure of $S$ even in the category II, 
if the system is located in the vicinity of the Kitaev's QSL phase.

To confirm the existence of the shoulder around $S=(Nk_{B}/2)\ln 2$ more quantitatively,
we decompose the temperature dependence of $S/Nk_{B}\ln 2$ by employing a phenomenological fitting function,
\eqsa{
\sigma (T)
= 
\sum_{\ell=1,2}\tau_{\ell}(T),
\label{fit_sigma}
}
where $\tau_{\ell}(T)$ is defined as
\cyan{
\begin{eqnarray}
\tau_{\ell}(T)=
\frac{\rho_{\ell}/2}{\displaystyle 1+\exp \left[\left(\frac{\beta_{\ell}+\gamma_{\ell}T_{0\ell}/T}{1+T_{0\ell}/T}\right)\ln\left(\frac{T_{0\ell}}{T}\right)\right]},
\label{fit_tau}
\end{eqnarray}
with \textcolor{black}{constraints $T_{01}\geq T_{02}$ and $\rho_1 + \rho_2 = 2$}.}
The specific functions $\tau_{\ell}(T)$ are chosen to describe power-low asymptotic behaviors
at both of low and high temperature limits, \cyan{as detailed in Appendix \ref{Appendix_Dec_ent}.}
As shown in \tbr{Fig.~\ref{fig_decomp_phi}}, we successfully decompose $S$
\cyan{of $\hat{H}_{\rm CJK}$ at $\varphi=90^{\circ}$} into two components.
\cyan{Here, we employ the standard least square fitting of the expectation values of $S/Nk_{B}\ln 2$ for $N=32$, with the fitting function $\sigma (T)$.} 
The ratio of the weight for two
\cyan{parts, $\tau_{\ell}(T)$,}
satisfies ${\rho_1}/{\rho_2} \sim 1$ in the Kitaev's QSL phase \cyan{at $\varphi=90^{\circ}$}, as expected, and even in the category II.
When the system goes toward deep inside of the magnetic ordered phase of the category I, 
the shoulder in $S$ continuously varies and finally disappears.
From \cyan{Fig.~\ref{fig_rho_phi},} 
we confirm that the ratio ${\rho_1}/{\rho_2}$ also \cyan{largely deviates from 1} in the category I.
Therefore, the successful fitting \cyan{with ${\rho_1}/{\rho_2} \sim 1$} in the category II supports that
the system is located in the vicinity of the Kitaev's QSL phase.
\textcolor{black}{Here, we note that, in Fig.~\ref{fig_rho_phi},
due to $2S+1$-fold ground-state degeneracy in the ferromagnetic phase at $\varphi=180^{\circ}$
with the total spin $S_{\rm tot}=N/2$,
$\rho_1$ and $\rho_2$ are rescaled by a factor $\left[1-(N\ln 2)^{-1}\ln (2S_{\rm tot}+1)\right]^{-1}$
with $N=32$ and the total spin $S_{\rm tot}=N/2=16$.}

We apply the same analysis to the {\it ab initio} Hamiltonian case.
In Fig.\ref{fig_S_T_ablambda}, we show temperature dependences of $S$. 
The two-peak structure of $C$ is common for the entire parameter range, $0\leq \lambda \leq 1$, and 
the plateau or shoulder around $S=(Nk_B/2)\ln 2$ is always observed.
At least, for $\lambda=0$, the value $(Nk_B/2)\ln 2$ corresponds to 
the contribution of the free Majorana fermions obviously, because the system 
is in the Kitaev's QSL phase.

As shown in \tbr{Fig.\ref{fig_decomp_ab}}, 
we also successfully decompose $S$ into two components \cyan{for the {\it ab initio} model, $\lambda=1$.}
The obtained \cyan{fitting parameters} are 
\cyan{$\beta_1=1.74\pm0.04$, $\gamma_1=1.88\pm0.09$, $T_{01}=120\pm 2$ K,
$\beta_2=2.26\pm0.06$, $\gamma_2=2.64\pm0.01$, $T_{02}=13.6\pm0.1$ K, and $\rho_1=0.99\pm 0.02$ with
the constraint $\rho_2=2-\rho_1$}.
This successful fitting also supports that
the shoulder structure in the temperature dependences of $S$
is a hallmark that Na$_2$IrO$_3$ is located in the vicinity of the Kitaev's QSL phase.
\cyan{As expected from the successful fitting for $\lambda=1$,
the decomposition is also successful for $0\leq\lambda < 1$ with $\rho_1/\rho_2 \sim 1$, as shown in Fig.\ref{fig_rho_ab}.}

After confirming the shoulder structure around $S=(Nk_B/2)\ln 2$,
a quantitative measure for distance between the target system in the category II and the Kitaev's QSL
can be introduced: The ratio of $T_{\ell}$ to $T_{h}$ gives a quantitative
measure for the distance, where $T_{\ell}$ ($T_{h}$) is the location of
the low-temperature (high-temperature) peak of the specific heat $C$.
\cyan{Details in the estimation of $T_{\ell}$ and $T_{h}$ are given in Appendix \ref{Appendix_ratio}.}
The two temperature scales $T_{01}$ and $T_{02}$ \tm{used in Eq.(\ref{fit_tau})} 
\cyan{\tm{roughly agree with} }
$T_{h}$ and $T_{\ell}$, respectively.
From the results of $C$ for $N=32$ shown in Fig.\ref{fig_TdepC_ABINITIO},
the ratio $T_{\ell}/T_{h}$ is smaller than \cyan{$0.03$} for the Kitaev's QSLs.
By taking into account the $N$-dependence of $T_{\ell}$, namely, the monotonic decreases in $T_{\ell}$
as $N$ increases, in the category III,
the condition \cyan{$T_{\ell}/T_{h}= 0.03$} gives the upper limit on the ratio $T_{\ell}/T_{h}$
of the Kitaev's QSLs.

In Fig.\ref{fig_phase_diagramv2}, the ratio \cyan{$T_{\ell}/T_{h}$} obtained through the fitting is summarized.
As shown in Fig.~\ref{fig_TdepC_ABINITIO}, for $\lambda=1$, the $N$-dependence is almost converged up to $N=32$.
Therefore, the estimate on the ratio
\cyan{$T_{\ell}/T_{h}\sim 0.11$}
offers a prediction for experimental observations,
which simultaneously offers experimental test on the {\it ab initio} Hamiltonian for Na$_2$IrO$_3$.
\tm{In the} inset of Fig.\ref{fig_phase_diagramv2}, the schematic phase diagram expected at the thermodynamic limit is shown.
From development of spin correlations shown in Fig.\ref{fig_abSS24}, the transition temperatures $T_{N}$
below which the zigzag orders set in are expected just below the low temperature scale \cyan{$T_{\ell}\sim T_{02}$}.
Note that the present two-dimensional {\it ab initio} Hamiltonian has \tm{a} magnetic anisotropy.
Thus, it is expected to show \tm{a} finite-temperature spontaneous symmetry breaking \tm{in} the thermodynamic limit.


\section{Discussion}

\subsection{Isoelectric doping and new materials}
\tbr{There exist experimental attempts to realize the Kitaev's spin liquid by isoelectronic doping
starting with Na$_2$IrO$_3$.
However, these attempts are not so successful so far, as explained below.}

\tbr{Besides a search for new materials such as Li$_2$RhO$_3$~\cite{PhysRevB.87.161121,PhysRevB.88.035115} and
$\alpha$-RuCl$_3$~\cite{PhysRevB.90.041112,PhysRevB.91.144420,PhysRevB.91.094422},
(Na$_{1-x}$Li$_x$)$_2$IrO$_3$
interpolating Na$_2$IrO$_3$ and
Li$_2$IrO$_3$ has been studied~\cite{PhysRevB.88.220414,PhysRevB.89.245113}.
In Ref.\onlinecite{PhysRevB.88.220414},
the N\'eel temperature $T_{N}$ is reported to be minimized around $x\sim 0.7$ and, simultaneously,
the frustration parameter defined as the ratio of the Curie-Weiss constant $\Theta$
and $T_{N}$, $\Theta/T_{N}$, becomes maximum.
The low temperature scale $T_{\ell}$ may also show minimum around $x\sim 0.7$,
which seemingly suggests that (Na$_{0.3}$Li$_{0.7}$)$_2$IrO$_3$ is a good candidate of the Kitaev's spin liquid.}

\tbr{However, we should note that there are several remaining issues in (Na$_{1-x}$Li$_x$)$_2$IrO$_3$
as a hunting field of the Kitaev's spin liquid.
First of all, phase separation for, at least, $0.25\lesssim x \lesssim 0.6$ is reported and
stability around $x\sim 0.7$ has not been confirmed yet.
Second, the reported specific heat coefficient $C/T$ around the low-temperature peaks
seems too small around $x\sim 0.7$ to exhaust $(Nk_{B}/2)\ln 2$.
An excess entropy release that reduces $C/T$ at low temperatures may be attributed to distortion due to the isoelectronic doping.}
\textcolor{black}{Our {\it ab initio} studies suggest that (Na$_{1-x}$Li$_x$)$_2$IrO$_3$
has a smaller lattice constant for larger $x$ because of smaller ionic radius of Li.
This may enhance the further neighbor transfers that are harmful for realizing the Kitaev's spin liquid.}

\subsection{Thin films}
\tbr{Making thin films of Na$_2$IrO$_3$ and related materials
on various substrates is another unexplored but promising approach to realize the Kitaev's spin liquid.
For example,
making thin films
of a perovskite iridate CaIrO$_3$ is efficient to stabilize the perovskite crystal structure
unstable as a bulk crystal~\cite{Hirai2015}, and is demonstrated to change the lattice constant depending on
the substrates.
\textcolor{black}{Thin films of Na$_2$IrO$_3$ on an appropriate substrate may expand the lattice constant
in comparison to the bulk Na$_2$IrO$_3$, which may decrease the other exchange couplings
relative to the Kitaev exchange.
This effectively reduces $\lambda$ and would stabilize the Kitaev's spin liquid.}
A combinatorial specific-heat measurement over the thin films of Na$_2$IrO$_3$ and related materials on various substrates
is highly desirable although specific-heat measurements of thin films that need sophisticated micocalorimeters~\cite{RevModPhys.78.217} are
not easy to carry out.
\textcolor{black}{Our criteria for the closeness to the Kitaev's spin liquid may help and
point to the favorable direction of efforts.}}


\section{Summary}

We have studied the magnetic excitations and specific heat 
of the generalized Kitaev-Heisenberg model and
the {\it ab initio} effective Hamiltonian of ${\rm Na_2IrO_3}$.
By comparing the linear spin wave dispersion, the dynamical spin structure factors,
temperature dependences of the specific heat,
we found that the parameter space of the effective Hamiltonians for ${\rm Na_2IrO_3}$ can be
\tm{classified to} three categories;
the phase diagram of the generalized Kitaev-Heisenberg model
exhibits two qualitatively distinct regions
in the magnetically ordered phase
in addition to the Kitaev's spin liquid phase.

\tm{In one region of the magnetically ordered phase,}
\tm{the specific heat has two-peak structure. In addition, } the conventional linear spin wave theory fails in explaining the low-lying excitation the dynamical spin structure factors and
the half-plateau-like temperature dependences of the entropy is observed owing to the thermal fractionalization of the spin degrees of freedom.
The other is the trivial region
\tm{located} far from the Kitaev's QSL.
In the trivial region, we observe the low-lying excitation well explained by linear spin wave theory and 
the single peak structure in the temperature dependence of the specific heat.

The {\it ab initio} Hamiltonian of ${\rm Na_2IrO_3}$, whose ground state is the zigzag magnetic order,
indeed shows the \tm{two peaks in the specific heat and indicates the} breakdown of the linear spin wave theory and the half-plateau-like temperature
dependences of the entropy pinned around $(Nk_B/2)\ln 2$,
which signals that the {\it ab initio} Hamiltonian of ${\rm Na_2IrO_3}$
is in the vicinity of the Kitaev's QSL phase.
These theoretical distinctions between the trivially ordered ground state and the system close to
the Kitaev's QSL,
offer experimental clues and criteria to understand a given material
in terms of the distance from the Kitaev's QSL phase.
It also offers a guideline for experiments to search the Kitaev materials.

\section*{Acknowledgments}
\label{ackno}
We thank T. Okubo, and T. Tohyama for fruitful discussions.  
This work was supported by 
\tm{the Computational Materials Science Initiative
(CMSI), and}
KAKENHI(Grants No. 25287104, 25287097, 15K05232, \tb{15K17702}, and No. 25287088) from MEXT Japan. 
We thank the computational resources of the K computer provided by the RIKEN Advanced Institute for Computational Science through the HPCI System Research project (hp120283, hp130081, \tm{hp140215 and hp150211}).
We also thank numerical resources in the ISSP Supercomputer Center at University of Tokyo  
and the Research Center for Nano-micro Structure Science and Engineering at University of Hyogo.

\appendix
\section{Decomposition of entropy $S(T)$}
\label{Appendix_Dec_ent}
\cyan{
Decomposition of $S(T)$ is examined to capture a sign of
the thermal fractionalization in the vicinity of the Kitaev spin liquids.
If the system is categorized into the category II or III,
$S(T)$ is decomposed into two parts with nearly equal weights.
Although the temperature dependence of $S(T)$ is complicated in general,
a simple ansatz defined below works as shown in \tbr{Fig.\ref{fig_decomp_phi} and Fig.\ref{fig_decomp_ab}}. }

\cyan{
To formulate our ansatz on the decomposition of $S(T)$,
we recall high-temperature and low-temperature behaviors of the entropy in simple systems.
As the simplest example of temperature dependence of entropy, the Schottky entropy,
\begin{eqnarray}
S_{\rm S}(T)
&=&\int_{0}^{T}dT'\frac{C_{\rm S}(T')}{T'}\nonumber\\
&=&\ln \left(1+e^{T_0/T}\right)-\frac{T_0}{T}\frac{e^{T_0/T}}{1+e^{T_0/T}},\label{ST}
\end{eqnarray}
with a characteristic temperature $T_0$ gives a typical high-temperature behavior as
$S_{\rm S}(T)\simeq \ln 2 - (T_0/T)^2 / 8$ for $T_0/T \ll 1$.
In addition to essentially singular behaviors in $S_{\rm S}(T)$ due to excitation gaps,
spin-wave-like power-law behaviors such as,
$S_{\rm S}(T)\propto T^2 + \mathcal{O}(T^3)$, are important at the low-temperature limit.
To capture the power-law behaviors, \tm{instead of Eq.(\ref{ST})}, we assume the following function to fit our numerical results of $S(T)$:
\begin{eqnarray}
\sigma (T)=\sum_{\ell=1,2}\tau_{\ell}(T),
\end{eqnarray}
where
\begin{eqnarray}
\tau_{\ell}(T)=
(\rho_{\ell}/2)/
\left[
1
+
\left(T_{0\ell}/T\right)^{p_{\ell}(T/T_{0\ell})}
\right],
\end{eqnarray}
with smooth functions $p_{\ell}(x)$.
To interpolate two power-law behaviors at high- and low-temperature limits and to mimic gap-like behaviors,
we employ one of the simplest rational form as
\begin{eqnarray}
p_{\ell}(T/T_{0\ell})=\frac{\beta_{\ell}+\gamma_{\ell}T_{0\ell}/T}{1+T_{0\ell}/T},
\end{eqnarray}
where $\beta_{\ell}$ and $\gamma_{\ell}$ correspond to exponents in the high- and low-temperature power-law behaviors, respectively.
}
\section{Ratio $T_{\ell}/T_{h}$}
\label{Appendix_ratio}
\cyan{In this Appendix, details \tm{are given for the prescription how to determine the temperature scales, $T_{\ell}$ and $T_{h}$, and their error bars.}}
\cyan{
The two temperature scales $T_{\ell}$ and $T_{h}$ of $\hat{H}_{\lambda}$
are simply determined as
\tm{those of} the peaks in
the specific heat $C(T)$ for the largest system size, $N=32$, in the present paper.
The higher temperature scale $T_{h}$ is determined with negligibly small uncertainty.
The lower temperature scale $T_{\ell}$ is, however, inevitably
\tm{under a numerical} uncertainty in $C(T)$, $\delta C(T)$, at low temperatures.
Thus, here, we estimate errors
by expanding $C(T)$ with respect to $T-T_{\ell}$ around $T=T_{\ell}$ as
$C(T)\simeq C(T_{\ell})-R(T-T_{\ell})^2$.
Then, \tm{the} uncertainty in $T_{\ell}$ is naturally estimated as
$\delta T_{\ell}=\left(\overline{\delta C}/R\right)^{1/2}$
with the upper bound of $\delta C(T)$, $\overline{\delta C} =\max_{T}\left\{ \delta C(T) \right\}$.
The ratio of the temperature scales $T_{\ell}/T_{h}$ is shown in Fig.\ref{fig_phase_diagramv2}
with the error bars.
}


\bibliography{Na2IrO3_finiteT.bib} 

\end{document}